%% file: paper.tex
\def\doublecol{1}

\if
\documentclass[journal]{IEEEtran}
\else
\documentclass[journal, onecolumn]{IEEEtran}
\fi

\input{header}
\usepackage{cite}
\usepackage{CJKutf8}

\begin{document}
  \title{Variational Secret Common Randomness Extraction}

  \author{Xinyang~Li \orcidlink{0000-0001-7262-5948},~\IEEEmembership{Graduate Student Member,~IEEE,}
  Vlad~C.~Andrei \orcidlink{0000-0001-5443-0100},~\IEEEmembership{Graduate Student Member,~IEEE,}
  Peter~J.~Gu~\orcidlink{0009-0002-7853-9201},~\IEEEmembership{Graduate Student Member,~IEEE,}
  Yiqi~Chen \orcidlink{0000-0002-4850-2072},~\IEEEmembership{Member,~IEEE,}\\
  Ullrich~J.~M\"onich \orcidlink{0000-0002-2390-7524},~\IEEEmembership{Senior Member,~IEEE,}
  and~Holger~Boche \orcidlink{0000-0002-8375-8946},~\IEEEmembership{Fellow,~IEEE}%
  \thanks{The authors are with the Department of Electrical and Computer
  Engineering, Technical University of Munich, Munich, 80333 Germany (e-mail: \{xinyang.li,
  vlad.andrei, peter.gu, yiqi.chen, moenich, boche\}@tum.de).}%
  \thanks{This work was supported by the Bavarian Ministry of Economic Affairs, Regional
  Development and Energy within the project ``6G Future Lab Bavaria''; by the project
  ``Next Generation AI Computing (gAIn)'', funded by the Bavarian Ministry of Science and
  the Arts and the Saxon Ministry for Science, Culture and Tourism; and by the Federal
  Ministry of Research, Technology and Space (BMFTR) under the programme CommUnity
  (16KISS013) and the programme ``Souver\"an. Digital. Vernetzt.'', joint project 6G-life
  (16KIS2414).}%
  \thanks{The code is available at \url{https://github.com/xinyanglii/vcr}.}
  }

    \maketitle

    \makeatletter
    \def\ps@IEEEtitlepagestyle{%
      \def\@oddfoot{\mycopyrightnotice}%
      \def\@oddhead{\hbox{}\@IEEEheaderstyle\leftmark\hfil\thepage}\relax
      \def\@evenhead{\@IEEEheaderstyle\thepage\hfil\leftmark\hbox{}}\relax
      \def\@evenfoot{}%
    }
    \def\mycopyrightnotice{%
      \begin{minipage}{\textwidth}
      \centering \scriptsize
      \end{minipage}
    }
    \makeatother

  \begin{abstract}

    Extracting \ac{cr} or secret keys from correlated sources observed by two legitimate parties in the presence of an eavesdropper is a classical information-theoretic problem with direct applications to physical-layer security. We propose a two-stage framework in which Alice and Bob first learn probabilistic neural encoders, termed \ac{vpq}, that jointly steer their discrete outputs toward near-uniformity, high agreement, and low leakage via a variational adversarial objective, and then reconcile the outputs into an identical key through a code-offset secure sketch. Because the adversarial objective suppresses the leakage already at the quantization stage, the reconciled key leaks no more than the quantizer outputs do, and no separate privacy-amplification step is needed. On the theoretical side, we prove that the \ac{vpq} training losses lower-bound the one-way secret-key capacity of the induced discrete source. We further derive in closed form the optimal one-way key rate achievable uniformly over the class of \ac{iid} sources that share the \ac{vpq} output's alphabet size and mismatch probability, with uniform outputs and no leakage to Eve, and show that linear code-offset secure sketches attain this rate. As a representative application, we exploit the paired \ac{ra} maps jointly sensed by Alice and Bob in \ac{isac} systems as the \ac{cr} source, making the scheme Alice-transmit-only and well suited to high-mobility settings. The framework is benchmarked against one classical and two recent learning-based methods on synthesized fading channels, validated end to end on \ac{ra} maps synthesized from 5G~NR simulations, and evaluated at the quantization stage on real-world \ac{sdr} measurements in a furnished lab room and an anechoic chamber. Robustness is examined across a grid of passive correlated eavesdroppers with controllable position uncertainty, down to a perfect-position attacker who observes Bob's exact range and angle.
  \end{abstract}

  \begin{IEEEkeywords}
    Common randomness, variational learning, physical layer security, integrated sensing and communications, secret key generation.
  \end{IEEEkeywords}

  \IEEEpeerreviewmaketitle

  \glsresetall

  \section{Introduction}
  \label{sec:intro}
  \input{paper-sections/intro}

  \section{Secret Common Randomness}\label{sec:cr}
  \input{paper-sections/cr}

  \section{Proposed Method}
  \input{paper-sections/method}

  \section{Sensing-Based Physical Layer Key Generation}
  \input{paper-sections/s4pls}

  \section{Conclusion}
  \input{paper-sections/conclusion}

  \section*{Acknowledgment}

  The device illustrations in Fig.~\ref{fig:hw}(a) were generated with OpenAI GPT Image (via Codex). The authors drew the signal flow and the labels.

  \ifCLASSOPTIONcaptionsoff
  \newpage
  \fi

  \bibliographystyle{IEEEtran}
  \bibliography{IEEEabrv,mybib}

  \begin{IEEEbiography}[{\includegraphics[width=1in,height=1.25in,clip,keepaspectratio]{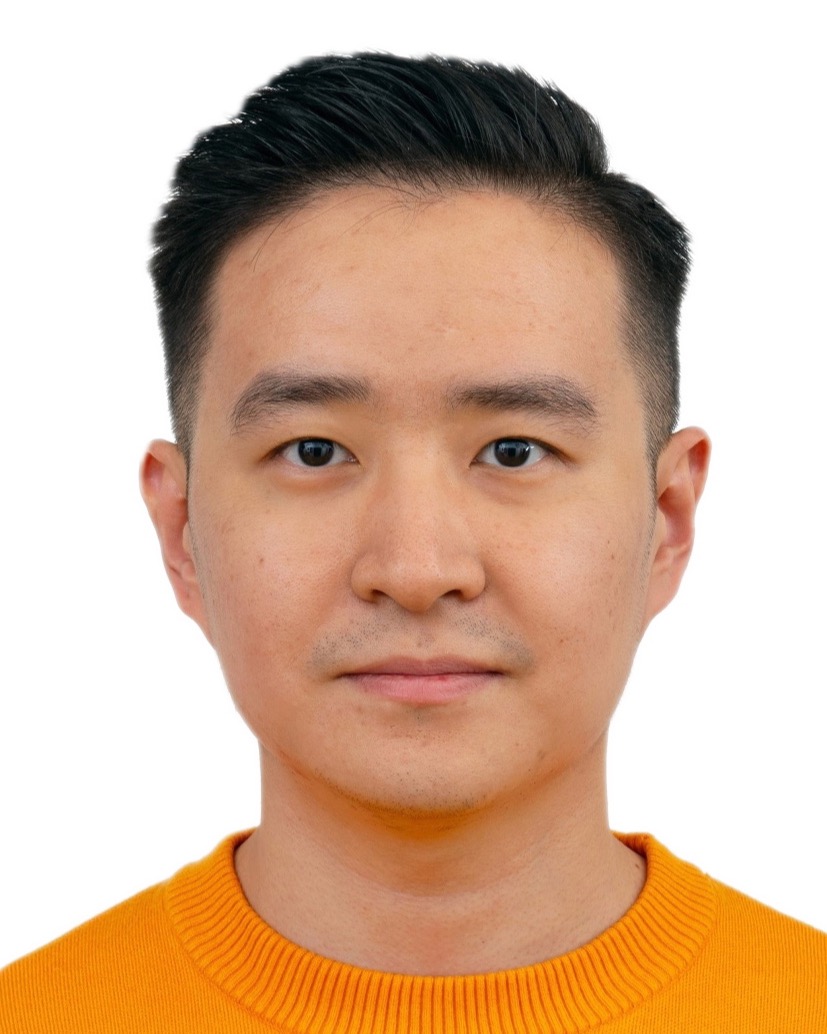}}]{Xinyang Li}
  (Graduate Student Member, IEEE) received the B.Eng. degree in electronic science and technology from Xidian University, Xi'an, China, in 2018, and the M.Sc. degree in electrical engineering and information technology from the Technical University of Munich, Germany, in 2022. He is currently pursuing the Ph.D. degree with the Chair of Theoretical Information Technology, Technical University of Munich. His research interests include integrated sensing and communications, information theory, AI-RAN, and digital twins.
  \end{IEEEbiography}

  \begin{IEEEbiography}[{\includegraphics[width=1in,height=1.25in,clip,keepaspectratio]{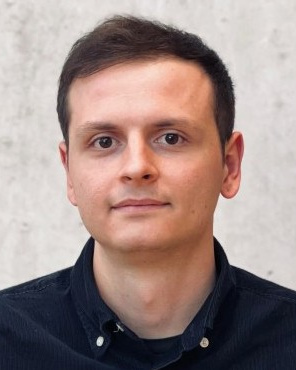}}]{Vlad C.\ Andrei}
  (Graduate Student Member, IEEE) received the M.Sc. degree from the Technical University of Munich, Munich, Germany, in 2021, where he is currently pursuing the Ph.D. degree with the Chair of Theoretical Information Technology under Prof.~Holger Boche. His research interests include 6G, integrated sensing and communication, wireless-aware robotic navigation, digital twins, agentic AI, and their hardware implementation. He received the Best Paper and Best Student Demo Awards at the IEEE Symposium on Joint Communications and Sensing in 2023 and 2024, respectively. He has served as a reviewer for multiple IEEE conferences, as well as for IEEE Transactions on Communications and IEEE Wireless Communications Letters, and as a program committee member of the IEEE Symposium on Joint Communications and Sensing.
  \end{IEEEbiography}

  \begin{IEEEbiography}[{\includegraphics[width=1in,height=1.25in,clip,keepaspectratio]{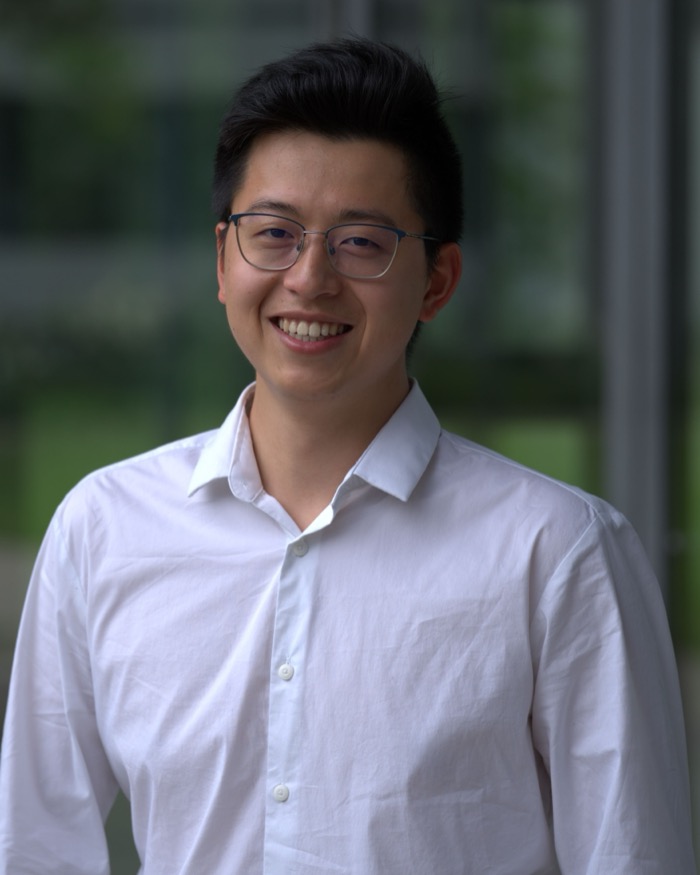}}]{Peter J.\ Gu}
  (Graduate Student Member, IEEE) received the B.Sc. and M.Sc. degrees in electrical engineering and information technology from the Karlsruhe Institute of Technology, Karlsruhe, Germany, in 2021 and 2023, respectively. Since 2024, he has been pursuing the Ph.D. degree with the Chair of Theoretical Information Technology, Technical University of Munich, Munich, Germany, where he is a member of the Advanced Communications and Embedded Security (ACES) Lab. His research interests include joint communication and sensing.
  \end{IEEEbiography}

  \begin{IEEEbiography}[{\includegraphics[width=1in,height=1.25in,clip,keepaspectratio]{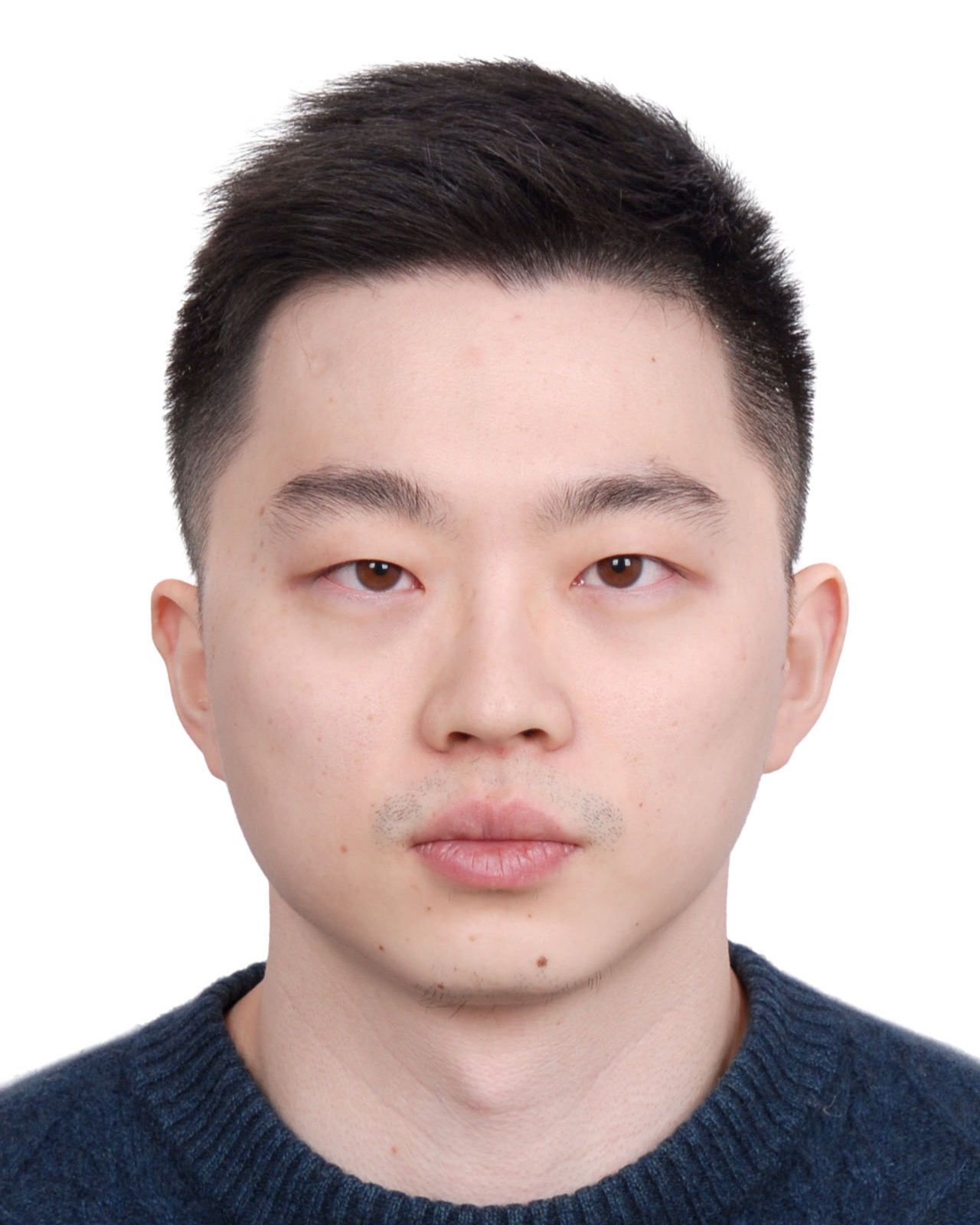}}]{Yiqi Chen}
  (Member, IEEE) received the B.E. degree in 2018 from Xidian University, Xi'an, China, and the Ph.D. degree in 2024 from Shanghai Jiao Tong University, Shanghai, China. He is currently a Postdoctoral Researcher with the Technical University of Munich, Munich, Germany. His research focuses on information theory, with a particular emphasis on Shannon theory, privacy, and security.
  \end{IEEEbiography}

  \begin{IEEEbiography}[{\includegraphics[width=1in,height=1.25in,clip,keepaspectratio]{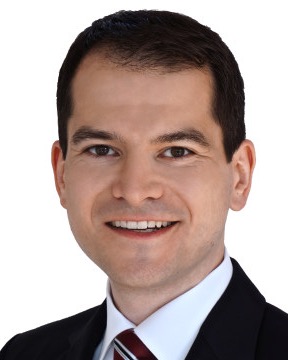}}]{Ullrich J.\ M{\"o}nich}
  (Senior Member, IEEE) received the Dipl.-Ing.\ degree in electrical engineering from the Technische Universit{\"a}t Berlin, Berlin, Germany, in 2005 and the Dr.-Ing.\ degree from the Technische Universit{\"a}t M{\"u}nchen, Munich, Germany, in 2011. During the winter term of 2003, he was a Visiting Researcher with the University of California, Santa Barbara. From 2005 to 2010, he was a Research and Teaching Assistant with the Technische Universit{\"a}t Berlin and from 2010 to 2012 with the Technische Universit{\"a}t M{\"u}nchen. From 2012 to 2015, he was a Postdoctoral Fellow with the Massachusetts Institute of Technology. Since 2015, he has been a Senior Researcher and Lecturer with the Technische Universit{\"a}t M{\"u}nchen, and since 2019, he has led the research activities in the Advanced Communication Systems and Embedded Security Lab (ACES Lab) at the Chair of Theoretical Information Technology. His research activities comprise signal and sampling theory, mobile communications, physical layer security, applied functional and harmonic analysis, and forensic DNA signal analysis. He received the Rohde \& Schwarz Award for his dissertation in 2012.
  \end{IEEEbiography}

  \begin{IEEEbiography}[{\includegraphics[width=1in,height=1.25in,clip,keepaspectratio]{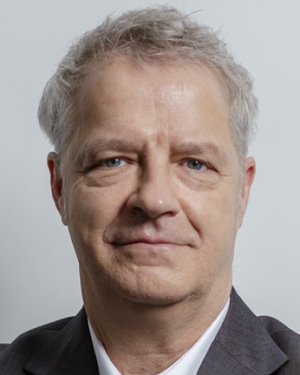}}]{Holger Boche}
  (Fellow, IEEE) received the Dipl.-Ing. degree in electrical engineering, the Dipl.-Math. degree in mathematics, and the Dr.-Ing. degree in electrical engineering from the Technische Universit{\"a}t Dresden, Germany, in 1990, 1992, and 1994, respectively, and the Dr. rer. nat. degree in pure mathematics from the Technische Universit{\"a}t Berlin, Germany, in 1998. He is currently a Full Professor with the Chair of Theoretical Information Technology, Technische Universit{\"a}t M{\"u}nchen, Germany, where he joined in 2010. Since 2021, he has been leading jointly with Frank Fitzek and Wolfgang Kellerer the BMFTR Research Hub 6G-life. He was elected as a member of the German Academy of Sciences (Leopoldina) in 2008 and Berlin Brandenburg Academy of Sciences and Humanities in 2009. He received the ``Innovation Award'' from the Vodafone Foundation in 2006 and the Gottfried Wilhelm Leibniz Prize from the German Research Foundation in 2008.
  \end{IEEEbiography}
\end{document}

%% file: header.tex
\usepackage{
  amsmath,
  bm,
  amssymb,
  algorithm,
  algpseudocode,
  mathtools,
  svg,
  ifluatex,
  pgfplots,
  amsthm,
  caption,
  subcaption,
  bookmark,
  hyperref,
  cite,
  multirow,
  booktabs,
  url,
  tikz,
  bbm,
  orcidlink,
  enumitem,
  soul,
  siunitx
}
\usetikzlibrary{graphs, positioning, quotes, shapes.geometric, arrows.meta, fit, calc, backgrounds}
\usepackage[acronym, shortcuts]{glossaries}
\usepackage{ifthen}
\usepackage{placeins}
\allowdisplaybreaks

\pgfplotsset{compat=1.15}
\renewcommand{\arraystretch}{1.7}

\newcommand{\brsq}[1]{\left[ #1 \right]}
\newcommand{\brcur}[1]{\left\{ #1 \right\}}

\newcommand{\RR}{\mathbb{R}}

\newcommand{\herm}{\mathsf{H}}
\newcommand{\nifft}{N_\mathrm{IFFT}}

\newcommand{\expc}[2]{\mathbb{E}_{#1}\brsq{#2}}
\newcommand{\expcs}[2]{\mathbb{E}_{#1}[#2]}

\newcommand{\pr}[1]{\mathrm{Pr}\brcur{#1}}

\newcommand{\ma}{\mathrm{A}}
\newcommand{\mb}{\mathrm{B}}
\newcommand{\me}{\mathrm{E}}
\newcommand{\mr}{\mathrm{Rx}}
\newcommand{\mt}{\mathrm{Tx}}
\newcommand{\rcs}{\sigma_{\mathrm{RCS}}}

\newcommand{\bX}{\bm{X}}

\newcommand{\nr}{N_r}
\newcommand{\na}{N_a}

\newcommand{\calL}{\mathcal{L}}
\newcommand{\calC}{\mathcal{C}}

\newcommand{\calN}{\mathcal{N}}

\newcommand{\calW}{\mathcal{W}}
\newcommand{\calM}{\mathcal{M}}

\newcommand{\calK}{\mathcal{K}}
\newcommand{\calI}{\mathcal{I}}
\newcommand{\calF}{\mathcal{F}}

\newcommand{\nsub}{n_{\mathrm{sc}}}
\newcommand{\nsym}{n_{\mathrm{sym}}}

\newcommand{\GF}{\mathrm{GF}}
\newcommand{\RS}{\mathrm{RS}}

\newacronym{isac}{ISAC}{integrated sensing and communications} \newacronym{los}{LoS}{line of sight}
\newacronym{sdr}{SDR}{software-defined radio} \newacronym{ofdm}{OFDM}{orthogonal frequency-division multiplexing}
\newacronym{aoa}{AoA}{angle of arrival} \newacronym{aod}{AoD}{angle of departure}
\newacronym{rcs}{RCS}{radar cross section} \newacronym{awgn}{AWGN}{additive white Gaussian noise}
\newacronym{cfo}{CFO}{carrier frequency offset} \newacronym{psk}{PSK}{phase shift keying}
\newacronym{cp}{CP}{cyclic prefix} \newacronym{ifft}{IFFT}{inverse fast Fourier transform}
\newacronym{ra}{RA}{range--angle} \newacronym{pls}{PLS}{physical layer security} \newacronym{plk}{PLK}{physical layer key}
\newacronym{cr}{CR}{common randomness}
\newacronym{sk}{SK}{secret key} \newacronym{nn}{NN}{neural network} \newacronym{fcn}{FCN}{fully-connected network}
\newacronym{snr}{SNR}{signal-to-noise ratio} \newacronym{cdf}{CDF}{cumulative distribution function} \newacronym{kmr}{KMR}{key mismatch rate} \newacronym{rs}{RS}{Reed--Solomon} \newacronym{vlb}{VLB}{variational lower bound} \newacronym{vub}{VUB}{variational upper bound}
\newacronym{rv}{RV}{random variable} \newacronym{kld}{KLD}{Kullback–Leibler divergence}
\newacronym{ema}{EMA}{exponential moving average} \newacronym{iid}{i.i.d.}{independent and identically distributed}
\newacronym{gkw}{GKW}{G{\'a}cs--K{\"o}rner--Witsenhausen} \newacronym{cdl}{CDL}{clustered delay line}
\newacronym{pdsch}{PDSCH}{physical downlink shared channel} \newacronym{iic}{IIC}{invariant information clustering}
\newacronym{aces}{ACES}{Advanced Communication Systems and Embedded Security}
\newacronym{tum}{TUM}{Technical University of Munich} \newacronym{usrp}{USRP}{universal software radio peripheral}
\newacronym{if}{IF}{intermediate frequency}
\newacronym{vpq}{VPQ}{variational probabilistic quantization}

\newtheorem{theorem}{Theorem}

\newtheorem{corollary}{Corollary}
\newtheorem{prop}{Proposition}

\theoremstyle{definition}
\newtheorem{definition}{Definition}

\newtheorem{remark}{Remark}

\glsdisablehyper \allowdisplaybreaks

\tikzset{%
block-common/.style={draw, fill=white, minimum height=1.5em, minimum width=4em}, block/.style={rectangle, block-common}, smallblock/.style={draw, fill=white, minimum height=1em, minimum width=1em}, bigblock/.style={block, minimum height=8em}, txtblock/.style={block, align=center, minimum height=4em}, txtbigblock/.style={bigblock, align=center}, input/.style={inner sep=1pt}, output/.style={inner sep=1pt}, sum/.style = {draw, fill=white, circle, minimum size=1.1em, inner sep=0pt, font={\small$+$}}, prod/.style = {draw, fill=white, circle, minimum size=1.1em, inner sep=0pt, font={\normalsize$\times$}}, pinstyle/.style = {pin edge={to-,thin,black}} }

%% file: paper-sections/intro.tex
\subsection{Background and Related Works}

\Ac{cr}\cite{ahlswede2002common1, ahlswede2002common2} refers to identical \acp{rv} extracted by two parties, Alice and Bob, from correlated observations; when secrecy is required against an external observer Eve, the output must additionally be unpredictable from Eve's side information, and the concept underpins applications ranging from secure communications\cite{maurer1993secret} and identification codes\cite{ahlswede2002identification} to quantum cryptography\cite{portmann2022security}. The fundamental limit is the \ac{cr} capacity, which admits closed-form characterizations in terms of the \ac{gkw} common component~\cite{gacs1973common, witsenhausen1975sequences} without public discussion and in terms of an auxiliary-variable optimization~\cite{ahlswede2002common1, ahlswede2002common2} with one-way public discussion, together with extensions to multi-way and helper settings~\cite{el2011network, csiszar2002common}. These results are existential, however: they do not tell us how to extract \ac{cr} in practice when $p(x,y,z)$ is unknown and high-dimensional.

Most practical efforts focus on the narrower problem of \ac{plk} generation from wireless channel measurements\cite{nguyen2021security, ren2011secret, zeng2015physical}. The classical pipeline\cite{aono2005wireless, wang2011fast, zhang2016key} follows a fixed sequence of channel probing, hand-designed scalar quantization (typically a level-crossing or \ac{cdf} rule on a single observable), information reconciliation, and privacy amplification. Such quantizers adapt only to the marginal of a single scalar observable, ignore the joint structure of Alice's and Bob's observations, and treat secrecy through spatial-decorrelation assumptions with a downstream privacy-amplification step rather than by design.

Recent learning-based methods\cite{han2020physical, guo2025physical, zhou2023daae} replace the hand-designed quantizer with an autoencoder that maps Alice's and Bob's observations into reciprocal continuous latent features, followed by a separate, non-differentiable quantization step; secrecy is either deferred to downstream privacy amplification or, in \cite{zhou2023daae}, enforced by a domain-adversarial branch that decorrelates the legitimate latent from Eve's latent. Across these methods, however, the latent is continuous, discretization sits outside the training loop, and no objective is tied to an explicit information-theoretic key-rate bound.

These observations motivate us to map the correlated sources directly into the discrete key alphabet in a single end-to-end differentiable step, so that uniformity, agreement, and secrecy are controlled by the training objective itself rather than recovered post hoc. We propose \ac{vpq}: probabilistic \ac{nn} encoders that output categorical distributions over a finite alphabet and are jointly trained by a variational objective whose entropy, mismatch, and leakage terms are optimized adversarially against a predictor at Eve\cite{poole2019variational, cheng2020club}. A code-offset secure sketch\cite{dodis2004fuzzy} then reconciles the two encoder outputs into identical keys, and because \ac{vpq} bounds the leakage at the quantization stage, no separate privacy-amplification step is required. We establish two information-theoretic guarantees for this design: every improvement in the \ac{vpq} training losses corresponds to a provable improvement in an achievable one-way secret-key rate for the induced source, and for the reconciliation stage we give in closed form the optimal one-way key rate achievable uniformly over the class of \ac{iid} sources that share the \ac{vpq} output's alphabet size and mismatch probability, with uniform outputs and no leakage to Eve; this rate is attained by linear code-offset secure sketches, showing that the reconciliation stage is not the bottleneck of the two-stage design.

Classical \ac{plk} further relies on two-way channel probing for reciprocity, imposing overhead and degrading under mobility. \Ac{isac}\cite{liu2022integrated} offers an alternative: wireless nodes already sense their shared propagation environment while communicating, and the sensing data can itself serve as a \ac{cr} source. Existing \ac{isac}-security work is largely confined to waveform-level physical-layer security\cite{su2020secure,su2022secure,wang2024sensing,su2025integrating,wyner1975wire}, which breaks down when Eve's channel is stronger or her location unknown. As a direct instantiation of our framework, we therefore propose to use the paired \ac{ra} maps jointly sensed by Alice and Bob as the \ac{cr} source: a \ac{los} path plus a detectable echo from Bob makes the two maps strongly correlated, while Bob's mobility makes successive maps effectively independent between updates, and unlike reciprocity-based schemes only Alice needs to transmit.

\subsection{Contributions}

The main contributions of this work are summarized as follows:

\begin{itemize}
    \item We propose a two-stage \ac{cr} extraction framework combining a learning-based \ac{vpq} method with a code-offset secure sketch. \ac{vpq} employs probabilistic \ac{nn} encoders trained by a variational adversarial objective to map correlated observations into near-uniform, agreeing, and low-leakage discrete \acp{rv}.
    \item We give an information-theoretic analysis of the framework. Instantiating the one-way secret-key capacity formula of~\cite{maurer1993secret,ahlswede2002common1} at the \ac{vpq} output shows that the three training objectives jointly lower-bound the secret-key capacity of the induced discrete source. We further derive in closed form the optimal worst-case key rate over the induced source class: no scheme that is uniformly reliable and secure on the class can exceed it, and $q$-ary linear code-offset secure sketches attain it. We also prove a validity condition and a zero-leakage diagnostic for the variational upper bound under adversarial optimization.
    \item We evaluate the framework on synthesized correlated Gaussian sources as a canonical \ac{plk} scenario under no-Eve, uncorrelated-Eve, and correlated-Eve conditions, and benchmark it against one classical and two recent learning-based baselines across a wide \ac{snr} range, including the privacy amplification that the baselines require.
    \item We further propose a sensing-based \ac{plk} scheme instantiating the framework in \ac{isac} systems, using the paired \ac{ra} maps jointly sensed by Alice and Bob as the \ac{cr} source. The scheme is validated end to end on 5G~NR simulations and at the quantization stage on \ac{sdr} measurements, and stress-tested against passive position-aware eavesdroppers up to a perfect-position attacker.
\end{itemize}

\emph{Notation.} Random variables are written in upper case and their
realizations in lower case; $\bm x$ denotes a vector and $\bm X$ a matrix, and
the subscript $\ast\in\{\ma,\mb,\me\}$ indexes Alice, Bob and Eve. $H(\cdot)$
and $I(\cdot\,;\cdot)$ are the entropy and the mutual information in bits, and
$h_q(\cdot)$ is the $q$-ary entropy function. Alice and Bob quantize their
observations into $W$ and $V$ on a common alphabet $\calW$ of size
$q=|\calW|$, with mismatch rate $p_e=\Pr\{W\neq V\}$, and $\GF(q)$ is the
finite field of order $q$. In the sensing case study, $N_{\mathrm{sc}}$ is the
number of \ac{ofdm} subcarriers, $\nifft$ the \ac{ifft} length, and $\nr$ and
$\na$ the numbers of range bins and beam directions of a range--angle map.%

%% file: paper-sections/cr.tex
\begin{figure}[!t]
    \centering
    \includegraphics[width=\columnwidth]{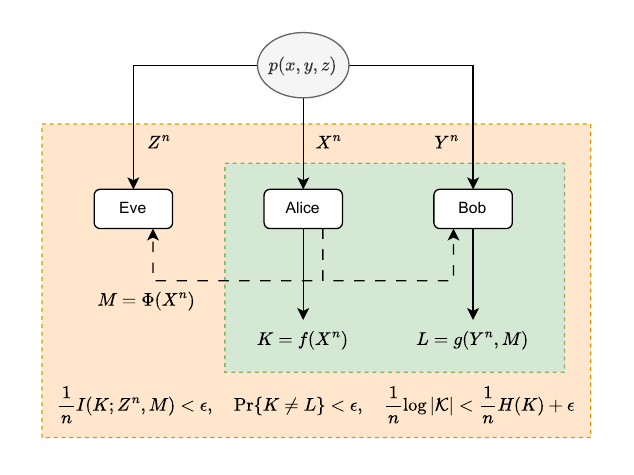}
    \caption{\ac{cr} extraction with one-way public discussion: Alice and Bob observe correlated sequences $X^n$ and $Y^n$ while Eve observes a correlated sequence $Z^n$; Alice sends a public message $M$ and both parties extract \acp{rv} $K,L$ that are required to agree, be nearly uniform, and be unpredictable from $(Z^n, M)$.}
    \label{fig:scr_blockdiagram}
\end{figure}

In many problems in information theory, \ac{cr} refers to \acp{rv} generated by two parties, Alice and Bob, from a pair of correlated random sources $(X,Y) \sim p(x,y)$ with the aid of public discussion\cite{ahlswede2002common1,ahlswede2002common2,ezzine2025uniform,ezzine2025compound}.
Specifically, let Alice observe a sequence $X^n = (X_1, \dots,X_n)$, while Bob observes $Y^n = (Y_1, \dots, Y_n)$. We consider the one-way communication setting, where Alice sends a public message $M = \Phi(X^n) \in \calM = \{1,\dots, |\calM|\}$ and both parties map their observations into \acp{rv}
\begin{equation}
    K = f(X^n), \quad L = g(Y^n, M),
\end{equation}
with $K, L \in \calK = \{1,\dots,|\calK|\}$, such that $K = L$ with high probability.
The mappings $f, g, \Phi$ may be either deterministic or stochastic.

An eavesdropper Eve is passive throughout this work: she observes a correlated sequence $Z^n$ jointly distributed with $(X^n, Y^n)$, together with the public message $M$. It is additionally desirable that the extracted \ac{cr} $K$ (or $L$) remains unpredictable from $(Z^n, M)$. This leads to the requirement that the averaged mutual information $\frac{1}{n} I(K; Z^n, M)$ is arbitrarily small, and $K$ tends to be uniformly distributed. The generated \acp{rv} $K$ or $L$ are also referred to as secret keys.
The entropy rate $\frac{1}{n} H(K)$ is called an achievable \ac{cr} rate and the supremum over all such achievable rates defines the \ac{cr} capacity. A schematic illustration of the one-way \ac{cr} extraction process is given in Fig.~\ref{fig:scr_blockdiagram}.
More formally, we have the following definition.

\begin{definition}\label{def:scr}
    Let Alice, Bob and Eve observe random sequences $\{(X_i, Y_i, Z_i)\}_{i=1}^n$ generated \ac{iid} from the joint distribution $p(x,y,z)$, with $Z=\varnothing$ if Eve is absent.
    A function $\Phi$ at Alice maps $X^n$ into a public message $M= \Phi(X^n) $, and
    a pair of functions $f,g$ extract \acp{rv} $K = f(X^n)$, $L=g(Y^n, M)$ at Alice and Bob, respectively. $K$ or $L$ is called \ac{cr} or secret key if the following conditions hold:
    \begin{align}
        &\pr{K \neq L} < \epsilon, \label{eq:key-agree-cond}\\
        &\frac{1}{n}\log|\calK| < \frac{1}{n}H(K) + \epsilon, \label{eq:key-entropy-cond} \\
        &\frac{1}{n}I(K; Z^n, M) < \epsilon.  \label{eq:key-leak-cond}
    \end{align}
    for every $\epsilon >0$ and sufficiently large $n$.
    The supremum of achievable entropy rates $\frac{1}{n}H(K)$ as $n\to \infty$ defines the \ac{cr} or secret key capacity.
\end{definition}

For the scenario without public discussion, i.e., $\Phi = \varnothing$,
it has been proved that the \ac{cr} capacity is given by $H(X_0|Z)$ where $X_0$ is the \ac{gkw} common component of $X$ and $Y$\cite{gacs1973common,witsenhausen1975sequences}. If $X,Y$ have an indecomposable joint distribution such as a joint Gaussian, the \ac{cr} capacity is zero. If Eve is absent, the \ac{cr} capacity becomes $I(X;Y)$ which is achieved by transmitting the compression of $X^n$ at rate $H(X|Y)$ such that Bob can decode $X^n$ losslessly with the side information $Y^n$ according to the Slepian--Wolf theorem\cite{slepian1973noiseless}. In the general one-way case with Eve present, the \ac{cr} capacity is
\begin{equation}\label{eq:sk-oneway}
    \max I(T; Y|U) - I(T; Z |U)
\end{equation}%

where the maximum is taken over all possible auxiliary \acp{rv} $(U,T)$ such that the Markov chain $U - T - X - (Y,Z)$ holds\cite{maurer1993secret,ahlswede2002common1}.

Although the theoretical properties of \ac{cr} have been well established, practical extraction methods remain scarce, especially when the joint distribution $p(x,y,z)$ is inaccessible and complicated. To this end, this work develops a two-stage \ac{cr} extraction framework that combines variational learning with secure-sketch reconciliation.

%% file: paper-sections/method.tex
\begin{figure*}[t]
    \centering
    \includegraphics{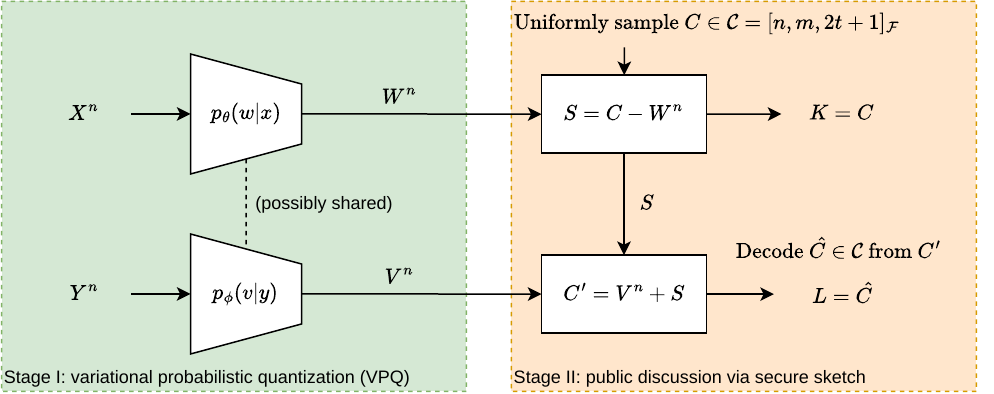}
    \caption{Overview of the proposed two-stage \ac{cr} extraction framework. In the first stage of \ac{vpq}, Alice and Bob map their respective observation sequences $X^n = (X_1, \dots, X_n)$, $Y^n = (Y_1,\dots, Y_n)$ into discrete, nearly uniform random sequences $W^n = (W_1, \dots, W_n)$, $V^n = (V_1,\dots, V_n)$ elementwise using \ac{nn} encoders. In the second stage, a secure sketch based on code-offset construction is applied for one-way public discussion.}
    \label{fig:method-overview}
\end{figure*}

The proposed \ac{cr} extraction framework has two stages. In the first stage, termed \ac{vpq}, Alice and Bob each employ a probabilistic \ac{nn} encoder, $p_\theta(w|x)$ and $p_\phi(v|y)$, mapping their observations into discrete \acp{rv} $(W,V)$ on a common finite alphabet $\calW$; training drives $(W,V)$ toward near-uniform marginals, a high agreement rate $\Pr\{W=V\}$, and small mutual information $I(W;Z)$ with Eve's side information $Z$. The last two layers of both encoders are a linear layer followed by a softmax of output dimension $|\calW|$, and the two encoders may share parameters when $X$ and $Y$ have the same structure. In the second stage, a secure sketch based on the code-offset construction reconciles the pair $(W^n,V^n)$ into an identical secret key $K=L$ over a one-way public channel. An overview is shown in Fig.~\ref{fig:method-overview}.

\subsection{Variational Probabilistic Quantization}\label{sec:vpq}

\emph{Mismatch rate.} The first objective minimizes $\Pr\{W\neq V\}$. Denoting the one-hot embeddings of $W,V$ by $\bm w,\bm v$, we have $\Pr\{W{=}V\mid x,y\}=\expcs{p_\theta(w|x)p_\phi(v|y)}{\bm w^\top \bm v}$, and the conditional independence of $W$ and $V$ given $(X,Y)$ factorizes this expectation as
\begin{align}
  \Pr\{W{=}V\mid x,y\}
  &= \expcs{p_\theta(w|x)}{\bm w}^{\top}\expcs{p_\phi(v|y)}{\bm v} \\
  &= \sum_{w'\in \calW} p_\theta(w'|x)\,p_\phi(w'|y),
\end{align}
so the empirical mismatch loss over a batch $\{(x_i,y_i)\}_{i=1}^{B}$ is
\begin{equation}
    \label{eq:lmr}
    \calL_{\mathrm{MR}} = -\frac{1}{B}\sum_{i=1}^{B}\sum_{w'\in \calW} p_\theta(w'|x_i)\,p_\phi(w'|y_i).
\end{equation}
Minimizing $\calL_{\mathrm{MR}}$ drives the two encoders toward a common (one-hot) output on each paired input.

\emph{Uniformity.} The second objective maximizes $H(W)$ and $H(V)$ toward $\log|\calW|$, which both prevents mode collapse (which $\calL_{\mathrm{MR}}$ alone would allow) and guarantees unpredictability. When $|\calW|$ exceeds the batch size, batch-wise marginal estimates are noisy, so we estimate the two marginals $p(w)$ and $p(v)$ by an \ac{ema} across training steps,
\begin{equation}
  p_{t}(w) = \alpha\, p_{t-1}(w) + \frac{1-\alpha}{B}\sum_{i=1}^{B} p_{\theta}(w|x_{i}),
\end{equation}
and analogously for $p_{t}(v)$, where $0\le\alpha<1$ and $p_{t-1}$ is detached from the gradient. The uniformity loss is
\begin{equation}
    \label{eq:lent}
    \calL_{\mathrm{ENT}} = -\frac{1}{2(1-\alpha)}\bigl(\hat{H}(W) + \hat{H}(V)\bigr),
\end{equation}
with $\hat{H}(W)=-\sum_{w\in\calW} p_{t}(w)\log p_{t}(w)$ and $\hat{H}(V)$ defined symmetrically; the factor $(1-\alpha)^{-1}$ rescales the current-batch gradient (the only term in the EMA recursion that is not detached) to match the un-EMA estimator.

\emph{Leakage rate.} The third objective suppresses $I(W;Z)$. When Eve is absent or uncorrelated, i.e.,\ $p(x,y,z)=p(x,y)p(z)$, the data-processing inequality gives $I(W;Z)\le I(X;Z)=0$ automatically, and the total loss reduces to
\begin{equation}
    \label{eq:lab}
    \calL_{\mathrm{AB}} = \calL_{\mathrm{ENT}} + \lambda_1 \calL_{\mathrm{MR}},\quad \lambda_1>0.
\end{equation}
When $Z$ is correlated with $(X,Y)$, $I(W;Z)$ is intractable without access to $p(w,z)$, so we optimize variational bounds adversarially against a learnable predictor $p_\psi(w|z)$ at Eve\cite{barber2004algorithm,poole2019variational,cheng2020club}. By the conditional independence $W\perp Z\mid X$, for any predictor $p_\psi$,
\begin{align}
    I(W;Z)
    &= \expc{p_\theta(w|x)p(x,z)}{\log p_\psi(w|z)} + H(W) \notag \\
    &\quad + \expcs{p(z)}{D(p(w|z)\|p_\psi(w|z))} \notag \\
    &\ge \expc{p_\theta(w|x)p(x,z)}{\log p_\psi(w|z)} + H(W) \notag \\
    &\triangleq I_{\mathrm{VLB}}(W;Z),
    \label{eq:ivlb-def}
\end{align}
with equality if and only if $p_\psi=p(w|z)$. The empirical estimator,
\begin{equation}
    \label{eq:vlb}
    \calI_{\mathrm{VLB}} = \frac{1}{B}\sum_{i=1}^{B}\sum_{w\in \calW} p_\theta(w|x_i)\log p_\psi(w|z_i),
\end{equation}
is the negative cross-entropy between $p_\theta$ and $p_\psi$, and is maximized over $\psi$ to train Eve's predictor to infer $W$ from $Z$ as tightly as possible. Minimizing $I_{\mathrm{VLB}}$ directly over $p_\theta$, however, both conflicts with entropy maximization and loosens the bound because updates to $p_\theta$ shift $p(w|z)$ away from the fixed $p_\psi$. We therefore adopt a variational upper bound obtained via Jensen's inequality,
\begin{align}
    I_{\mathrm{VUB}}(W;Z)
    &\triangleq \expc{p_\theta(w|x)p(x,z)}{\log p_\psi(w|z)} \notag \\
    &\quad - \expc{p(w)p(z)}{\log p_\psi(w|z)},
    \label{eq:ivub-def}
\end{align}
whose empirical estimator factorizes as
\begin{equation}
    \label{eq:vub}
    \calI_{\mathrm{VUB}} = \calI_{\mathrm{VLB}} - \frac{1}{B^{2}}\sum_{i,j=1}^{B}\sum_{w\in \calW} p_\theta(w|x_i)\log p_\psi(w|z_j).
\end{equation}
Minimizing $\calI_{\mathrm{VUB}}$ over $p_\theta$ therefore both reduces the predictor's accuracy and forces the encoder output to be approximately independent of $Z$. Training alternates: with $\theta,\phi$ fixed, $\psi$ is updated to maximize $\calI_{\mathrm{VLB}}$; with $\psi$ fixed, $\theta,\phi$ are updated to minimize $\calI_{\mathrm{VUB}}$. Under an approximate $p_\psi$, $\calI_{\mathrm{VUB}}$ is not always a true upper bound of $I(W;Z)$; Proposition~\ref{prop:var-gap} below makes the validity criterion and approximation error precise.

The gaps between the variational bounds and the true mutual
information $I(W;Z)$ admit exact KL-divergence characterizations\cite{barber2004algorithm,poole2019variational},
which yield a precise criterion for when $I_{\mathrm{VUB}}$ is a valid upper
bound, a controlled approximation-error guarantee when it is not, and a
model-free diagnostic for the zero-leakage regime.
\begin{prop}\label{prop:var-gap}
For any predictor $p_\psi(w \mid z)$, the gaps between the
variational bounds and the true mutual information are KL divergences:
\begin{align}
  I(W;Z) - I_{\mathrm{VLB}}(W;Z)
  &= \expcs{p(z)}{D\bigl(p(w|z) \| p_\psi(w|z)\bigr)},
    \label{eq:vlb-gap}\\
  I_{\mathrm{VUB}}(W;Z) - I_{\mathrm{VLB}}(W;Z)
  &= \expcs{p(z)}{D\bigl(p(w) \| p_\psi(w|z)\bigr)}.
    \label{eq:vub-vlb-gap}
\end{align}
Consequently, the VUB is a valid upper bound of $I(W;Z)$ if and only
if~\eqref{eq:vub-vlb-gap}$\,\ge\,$\eqref{eq:vlb-gap}, i.e.,
\begin{align}
  \expcs{p(z)}{D\bigl(p(w) \| p_\psi(w|z)\bigr)}
  \ge
  \expcs{p(z)}{D\bigl(p(w|z) \| p_\psi(w|z)\bigr)}.
    \label{eq:vub-valid}
\end{align}
Moreover, if $p_\psi(w|z)=p(w|z)$, then~\eqref{eq:vlb-gap} vanishes
and
\begin{align}
  I_{\mathrm{VUB}}(W;Z)
  = I(W;Z)
    + \expcs{p(z)}{D\bigl(p(w) \| p(w|z)\bigr)}.
    \label{eq:vub-overshoot}
\end{align}
\end{prop}

\begin{IEEEproof}
Conditioning the joint KL divergences $D(p(w,z)\,\|\,p_\psi(w|z)\,p(z))$ and $D(p(w)\,p(z)\,\|\,p_\psi(w|z)\,p(z))$ on $z$ and matching against the definitions in~\eqref{eq:ivlb-def} and~\eqref{eq:ivub-def} yields~\eqref{eq:vlb-gap} and~\eqref{eq:vub-vlb-gap}, respectively; the validity criterion~\eqref{eq:vub-valid} and the overshoot identity~\eqref{eq:vub-overshoot} follow immediately.
\end{IEEEproof}

Both gaps in~\eqref{eq:vlb-gap}--\eqref{eq:vub-vlb-gap} are nonnegative as averaged KL divergences, so $I_{\mathrm{VLB}} \le I(W;Z)$ always holds and~\eqref{eq:vub-vlb-gap} additionally implies $I_{\mathrm{VLB}}(W;Z) \le I_{\mathrm{VUB}}(W;Z)$ for every predictor $p_\psi$, independently of whether the VUB is a valid upper bound of $I(W;Z)$. The validity criterion~\eqref{eq:vub-valid} has a clear operational interpretation: the left-hand side measures how far $p_\psi(w|z)$ deviates from the marginal $p(w)$ (large when the predictor has learned to exploit $z$), while the right-hand side measures the posterior-fit error (small after the VLB-maximization step pushes $p_\psi$ toward $p(w|z)$).

\begin{corollary}\label{cor:zero-leak-vub}
If $I(W;Z)=0$, then $I_{\mathrm{VUB}}(W;Z) = 0$ for every predictor
$p_\psi(w|z)$.
\end{corollary}

\begin{IEEEproof}
If $I(W;Z)=0$, then $p(w,z)=p(w)\,p(z)$, so the two expectations in
the definition of $I_{\mathrm{VUB}}$ are identical.
\end{IEEEproof}

\begin{corollary}\label{cor:approx-error}
If the posterior-fit error satisfies
$\expcs{p(z)}{D(p(w|z)  \| p_\psi(w|z))} \le \varepsilon$,
then either $I_{\mathrm{VUB}}(W;Z) \ge I(W;Z)$ (VUB is valid),
or $0 < I(W;Z) - I_{\mathrm{VUB}}(W;Z) \le \varepsilon$.
\end{corollary}
\begin{IEEEproof}
By Proposition~\ref{prop:var-gap}, $I(W;Z) - I_{\mathrm{VUB}}(W;Z) = \text{RHS of}~\eqref{eq:vlb-gap} - \text{RHS of}~\eqref{eq:vub-vlb-gap}$. Let $G_1$ and $G_2$ denote the two RHSs; the assumption gives $G_1 \le \varepsilon$. If $G_2 \ge G_1$, then $I_{\mathrm{VUB}} \ge I(W;Z)$ and the VUB is valid. Otherwise $0 < G_1 - G_2 \le G_1 \le \varepsilon$.
\end{IEEEproof}

\begin{remark}\label{rem:lautum}
The non-negativity of~\eqref{eq:vub-vlb-gap} also yields a model-free diagnostic for zero leakage. When the predictor is well-trained so that $p_\psi(w|z)\approx p(w|z)$, the theoretical gap $I_{\mathrm{VUB}} - I_{\mathrm{VLB}}$ converges to the lautum information~\cite{palomar2008lautum} $L(W;Z) := \expcs{p(z)}{D(p(w) \| p(w|z))}$, which satisfies $L(W;Z) = 0$ if and only if $I(W;Z) = 0$. The empirical estimator $\calI_{\mathrm{VLB}}$ in~\eqref{eq:vlb} is the $H(W)$-free part of the theoretical $I_{\mathrm{VLB}}(W;Z) = \expc{p_\theta(w|x)p(x,z)}{\log p_\psi(w|z)} + H(W)$, so the empirical gap $\calI_{\mathrm{VUB}} - \calI_{\mathrm{VLB}}$ exceeds the theoretical gap $I_{\mathrm{VUB}} - I_{\mathrm{VLB}}$ by exactly $H(W)$ and converges to $H(W) + L(W;Z)$. Consequently, given a well-trained predictor, observing $\calI_{\mathrm{VUB}} - \calI_{\mathrm{VLB}} \to H(W)$ during training (which coincides with $\log|\calW|$ in the uniform-$W$ regime enforced by $\calL_{\mathrm{ENT}}$) is a sufficient empirical signature of the zero-leakage regime.
\end{remark}

\begin{algorithm}
    [t]
    \normalcolor
    \caption{\ac{vpq} Training Algorithm}
    \begin{algorithmic}
        \For{each training step $t=1,2,\dots$} \State Sample a data batch
        $\{(x_{i}, y_{i}, z_{i})\}_{i=1}^{B}$ \State \Comment{$z_{i}=\varnothing$ if Eve absent}
        \State Alice and Bob output $p_{\theta}(w|x_{i}), p_{\phi}(v|y_{i})$ for
        all $i$ \State Compute $\calL_{\mathrm{MR}}$ according to \eqref{eq:lmr}
        \State Compute $p_{t}(w)$ and $p_{t}(v)$ via \ac{ema} \State Compute $\calL
        _{\mathrm{ENT}}$ according to \eqref{eq:lent} \State
        $\calL_{\mathrm{AB}}\gets \calL_{\mathrm{ENT}} + \lambda_1\calL_{\mathrm{MR}}$
        \If{Eve is present} \If{update $\psi$} \State Eve outputs
        $p_{\psi}(w|z_{i})$ for all $i$ \State Compute $\calI_{\mathrm{VLB}}$
        according to \eqref{eq:vlb} \State Update $p_{\psi}$ by maximizing $\calI
        _{\mathrm{VLB}}$ \EndIf \If{update $\theta, \phi$} \State Eve outputs
        $p_{\psi}(w|z_{i})$ for all $i$ \State Compute $\calI_{\mathrm{VUB}}$
        according to \eqref{eq:vub}
        \State $\calL \gets \calL_{\mathrm{AB}}+ \lambda_2 \calI_{\mathrm{VUB}}$
        \State Update $p_{\theta}, p_{\phi}$ by minimizing $\calL$ \EndIf \Else
        \State Update $p_{\theta}, p_{\phi}$ by minimizing $\calL_{\mathrm{AB}}$
        \EndIf \EndFor
    \end{algorithmic}\label{alg:vske}
\end{algorithm}

\subsection{Achievable Key-Rate Analysis}\label{sec:key-rate-analysis}

The three \ac{vpq} objectives (entropy maximization, mismatch minimization,
and leakage suppression) admit a unified information-theoretic
interpretation.  We show below that they jointly lower-bound the one-way
secret-key capacity $C_{\mathrm{SK}}(W,V|Z)$ of the induced discrete source $(W,V,Z)$.

\begin{theorem}\label{thm:vpq-keyrate}
Let $W$ and $V$ take values in the same $q$-ary alphabet $\calW$ with
$q=|\calW|$, and let $p_e:=\Pr\{W\neq V\}$.  Then
\begin{equation}\label{eq:csk-bound}
  C_{\mathrm{SK}}(W,V|Z) \;\ge\; I(W;V)-I(W;Z).
\end{equation}
Consequently,
\begin{equation}\label{eq:csk-fano}
  C_{\mathrm{SK}}(W,V|Z)
  \;\ge\; H(W)-h_2(p_e)-p_e\log(q-1)-I(W;Z),
\end{equation}
where $h_2(\cdot)$ is the binary entropy function.
Moreover, if $I(W;Z)=0$, then
\begin{equation}\label{eq:csk-zero-leak}
  C_{\mathrm{SK}}(W,V|Z) = I(W;V).
\end{equation}%

\end{theorem}

\begin{IEEEproof}
In the one-way secret-key capacity formula recalled in Section~\ref{sec:cr} from~\cite{maurer1993secret,ahlswede2002common1}, choose $U\equiv\mathrm{const}$
and $T=W$.  The Markov constraint $U$--$T$--$W$--$(V,Z)$ is satisfied
trivially, yielding~\eqref{eq:csk-bound}.
For~\eqref{eq:csk-fano}, write $I(W;V)=H(W)-H(W|V)$ and apply Fano's
inequality with the estimator $\hat{W}(V)=V$, whose error probability is
$p_e$, giving $H(W|V)\le h_2(p_e)+p_e\log(q-1)$.
For~\eqref{eq:csk-zero-leak}, the lower bound
$C_{\mathrm{SK}}(W,V|Z)\ge I(W;V)$ is immediate
from~\eqref{eq:csk-bound} with $I(W;Z)=0$.
For the converse, $I(W;Z)=0$ together with the Markov chain $U$--$T$--$W$--$(V,Z)$ (which gives $(U,T)\perp Z\mid W$) yields $(U,T,W)\perp Z$, so $I(T;Z|U)=0$.
The capacity formula therefore reduces to
$C_{\mathrm{SK}}(W,V|Z)=\max I(T;V|U)$.
Since the Markov constraint now implies $U$--$T$--$W$--$V$, we have
$I(T;V|U)\le I(U,T;V)\le I(W;V)$,
where the last step follows from the data processing inequality applied to
the chain $(U,T)$--$W$--$V$.
Combining with the lower bound proves~\eqref{eq:csk-zero-leak}.%

\end{IEEEproof}

\begin{remark}\label{rem:aux}
The capacity in~\eqref{eq:sk-oneway} is defined by a maximization over auxiliary
\acp{rv} $(U,T)$. For the sources treated here, that maximization cannot be
evaluated: the joint law $p(x,y,z)$ is unknown, and $X$ and $Y$ are continuous
and high-dimensional. Computing such
quantities by optimizing over distributions has been studied in its own right,
including for continuous alphabets\cite{li2025capacity}.
Theorem~\ref{thm:vpq-keyrate} instead fixes $U\equiv\mathrm{const}$ and $T=W$,
which replaces the maximization by an optimization over the encoder parameters
$\theta$ by stochastic gradient descent, so the resulting statement is an
achievable bound rather than the capacity. The learned variable is categorical
on $\calW$, the alphabet of the code-offset sketch of
Section~\ref{sec:sketch}, so the auxiliary \ac{rv} of the rate expression and
the symbol alphabet of the reconciliation code coincide. \ac{vpq} thus gives a
constructive counterpart to the auxiliary \ac{rv}
of~\cite{maurer1993secret,ahlswede2002common1}, where it is an existence
statement.%

\end{remark}

Because the \ac{vpq} encoders define a stochastic mapping $(X,Y,Z)\mapsto(W,V,Z)$, any key-generation scheme for the induced source composes with the encoders to give a valid scheme for the original source, so $C_{\mathrm{SK}}(X,Y|Z)\ge C_{\mathrm{SK}}(W,V|Z)$. The bound~\eqref{eq:csk-fano} thus establishes a term-by-term correspondence between the \ac{vpq} objectives and the achievable key rate: $\calL_{\mathrm{ENT}}$ raises $H(W)$ toward $\log q$, $\calL_{\mathrm{MR}}$ shrinks the Fano penalty $h_2(p_e)+p_e\log(q-1)$, and adversarial minimization of $\calI_{\mathrm{VUB}}$ acts as a variational surrogate for $I(W;Z)$. Combining them, the overall \ac{vpq} loss
\begin{equation}
\calL = \calL_{\mathrm{AB}} + \lambda_2 \calI_{\mathrm{VUB}}
      = \calL_{\mathrm{ENT}} + \lambda_1\calL_{\mathrm{MR}} + \lambda_2\calI_{\mathrm{VUB}},
\end{equation}
with $\lambda_2\ge 0$, is a surrogate for maximizing the right-hand side of~\eqref{eq:csk-fano}; when Eve is absent, $I(W;Z)=0$ holds automatically by data processing and \eqref{eq:lab}'s loss $\calL_{\mathrm{AB}}$ directly maximizes the Fano lower bound on $I(W;V)$, which by~\eqref{eq:csk-zero-leak} equals the exact one-way secret-key capacity. In the adversarial case, to keep the two competing gradients comparable we update $\lambda_2$ following the scaling of VQ-GAN\cite{esser2021taming},
\begin{equation}
    \label{eq:lambda-update}
    \lambda_2 = \frac{\|\nabla_{\theta_L}\calL_{\mathrm{AB}}\|_{2}}{\|\nabla_{\theta_L}\calI_{\mathrm{VUB}}\|_{2}+\delta},
\end{equation}
where $\nabla_{\theta_L}$ denotes the gradient with respect to the last layer before the softmax of $p_\theta$ and $\delta=10^{-7}$ prevents division by zero. The resulting alternating optimization, summarized in Algorithm~\ref{alg:vske}, updates $\psi$ to maximize $\calI_{\mathrm{VLB}}$ with $\theta,\phi$ fixed, and $\theta,\phi$ to minimize $\calL$ with $\psi$ fixed.

\subsection{Secret Key Reconciliation}\label{sec:sketch}

Because the \ac{vpq} mismatch probability $p_e=\Pr\{W\neq V\}$ is generally nonzero, the agreement rate between $W^n$ and $V^n$ decays exponentially in $n$, so a public-discussion step is needed to turn the correlated pair into an identical key. We adopt the code-offset secure sketch\cite{dodis2004fuzzy} over the finite field $\calF=\GF(|\calW|)$ with an $[n,m,2t+1]_\calF$ error-correcting code $\calC$. Alice samples a codeword $C$ uniformly from $\calC$, forms
\begin{equation}
    S = C - W^n
\end{equation}
and sends $S$ over the public channel. Bob computes $C' = V^n + S = C + (V^n-W^n)$ and decodes $\hat C\in\calC$; when $d_H(V^n,W^n)\le t$, Bob recovers $\hat C=C$, and both parties set $K=C$, $L=\hat C$. Each reconciled block then consumes $n$ \ac{vpq} symbols and produces $m\log|\calW|$ key bits, giving the secret-key rate
\begin{equation}
    \frac{1}{n}H(K) = \frac{m}{n}\log|\calW|
\end{equation}
since $|\calC|=|\calW|^m$. Increasing $m$ raises the key rate and lowers the number of correctable errors.

We now analyze the fundamental limit of the key rate achievable by code-offset reconciliation.
Since only the mismatch probability $p_e$ and the uniformity of $W$ are controlled by the \ac{vpq} training, the appropriate benchmark is the worst-case universal key rate over all source laws compatible with these two quantities.
Define the $q$-ary entropy function $h_q(p):={({h_2(p)+p\log(q-1)})/{\log q}}$.

\begin{theorem}\label{thm:universal-rate}
Let $q=|\calW|$ be a prime power, and let $\mathcal{D}_{p_e}$ denote the class
of all \ac{iid} source triples $(W,V,Z)$ with $W,V$ uniform on $\GF(q)$,
$\Pr\{W\neq V\}=p_e<1-1/q$,\footnote{The constraint $p_e<1-1/q$ is imposed by the Hamming-ball decoder used in the achievability proof below; the Fano lower bound in Theorem~\ref{thm:vpq-keyrate} itself remains valid for all $p_e\in[0,1]$.} and $I(W;Z)=0$.
Then the optimal worst-case universal one-way key rate over $\mathcal{D}_{p_e}$ is
\begin{equation}\label{eq:universal-rate}
  R^{\star}_{\mathrm{univ}}
  = \log q - h_2(p_e) - p_e\log(q-1)
  = \bigl(1-h_q(p_e)\bigr)\log q
\end{equation}
bits per source symbol.
Every rate below $R^{\star}_{\mathrm{univ}}$ is achievable by a sequence of $q$-ary linear code-offset secure sketches $S_n=K_n-W^n$ under which Bob's decoded key $\hat{K}_n$ satisfies
$\Pr\{\hat{K}_n\neq K_n\}\to 0$
and $I(K_n;S_n,Z^n)=0$ for every~$n$, uniformly over $\mathcal{D}_{p_e}$.%

\end{theorem}

The achievability argument follows the standard random
linear coding construction over $\GF(q)$~\cite{csiszar1982linear}, here applied
to the source class $\mathcal{D}_{p_e}$, over which the rate must hold
uniformly.

\begin{IEEEproof}
\emph{Converse.}
Consider the source law in $\mathcal{D}_{p_e}$ where $Z$ is constant and
$P_{W|V}$ is the $q$-ary symmetric channel with crossover probability
$p_e/(q-1)$.
Then $I(W;V)=\log q - h_2(p_e)-p_e\log(q-1)$ and $I(W;Z)=0$, so
Theorem~\ref{thm:vpq-keyrate} gives
$C_{\mathrm{SK}}(W,V|Z)=I(W;V)=R^{\star}_{\mathrm{univ}}$.
Since this source law belongs to $\mathcal{D}_{p_e}$, any scheme that is
reliable and secure uniformly over $\mathcal{D}_{p_e}$ cannot achieve a key
rate exceeding $R^{\star}_{\mathrm{univ}}$.

\emph{Achievability.}
We follow the random-linear-code argument of~\cite{csiszar1982linear} for Slepian--Wolf coding.
Fix any target rate $R_{\mathrm{key}}<R^{\star}_{\mathrm{univ}}$ and
choose $\rho$ satisfying
$h_q(p_e)<\rho<(\log q - R_{\mathrm{key}})/\log q$.
By continuity of $h_q$, there exists $\eta>0$ such that
$p_e+\eta<1-1/q$ and $\rho>h_q(p_e+\eta)$.
Choose integers $m_n$ with $m_n/n\to\rho$, draw a random parity-check
matrix $H_n\in\GF(q)^{m_n\times n}$ with entries drawn i.i.d.\ uniformly from $\GF(q)$, and let
\begin{equation}
  T_n:=H_nW^{n\top}.
\end{equation}
Decode $W^n$ from the syndrome $T_n$ and the side information $V^n$
by searching for a unique $\hat{w}^n$ satisfying
$H_n\hat{w}^{n\top}=T_n$ and
$d_H(\hat{w}^n,V^n)\le n(p_e+\eta)$.
The error event is contained in the union of
$\{d_H(W^n,V^n)>n(p_e+\eta)\}$,
which vanishes by the law of large numbers, and the event that some
$\tilde{w}^n\neq W^n$ satisfies the same syndrome and lies in the same
Hamming ball.
For the latter, since the rows of $H_n$ are independent and each row annihilates a fixed nonzero vector with probability $1/q$, the syndrome of any fixed nonzero vector is uniform on $\GF(q)^{m_n}$, so the probability of a syndrome collision is $q^{-m_n}$.
By the union bound over the
\begin{equation}
  B_q(n,p_e{+}\eta)
  :=\sum_{i=0}^{\lfloor n(p_e+\eta)\rfloor}\binom{n}{i}(q-1)^i
\end{equation}
sequences in the Hamming ball\cite{csiszar2011information}, the collision probability is bounded by
$B_q(n,p_e{+}\eta)\,q^{-m_n}\to 0$,
since $\frac{1}{n}\log_q B_q(n,\alpha)\to h_q(\alpha)$ and
$m_n/n\to\rho>h_q(p_e{+}\eta)$.
For a linear code the collision event depends on $(W^n,V^n)$ only through
the error vector $e^n:=V^n-W^n$: decoding fails only if some
$\tilde{e}^n\neq e^n$ of Hamming weight at most $n(p_e{+}\eta)$ satisfies
$H_n(e^n-\tilde{e}^n)^\top=0$. Within each type class of $e^n$ the expected
fraction of vectors decoded incorrectly is at most
$B_q(n,p_e{+}\eta)\,q^{-m_n}$, so a union bound over the fewer than
$(n{+}1)^{q}$ types yields a deterministic sequence of parity-check matrices
whose error probability vanishes uniformly over
$\mathcal{D}_{p_e}$\cite{csiszar1982linear}.

Now let $\calC_{n}:=\ker(H_n)$, draw $K_n$ uniform on $\calC_{n}$
independently of $(W^n,V^n,Z^n)$, and define the code-offset sketch~\cite{dodis2004fuzzy}
$S_n:=K_n-W^n$.
Decoding $K_n$ from $(V^n,S_n)$ is equivalent to the syndrome-decoding
problem above: for any candidate $\hat{k}^n\in\calC_{n}$, setting
$\hat{w}^n:=\hat{k}^n-S_n$ gives $H_n\hat{w}^{n\top}=T_n$ and
$d_H(\hat{w}^n,V^n)=d_H(\hat{k}^n,V^n+S_n)$,
so $\Pr\{\hat{K}_n\neq K_n\}\to 0$ uniformly over $\mathcal{D}_{p_e}$.

\emph{Secrecy.}
Since $W^n$ is uniform on $\GF(q)^n$, $K_n$ is uniform on $\calC_{n}$, and
$K_n$ is independent of $W^n$, we have for every
$k\in\calC_{n}$ and $s\in\GF(q)^n$:
\begin{align}
  &P(K_n{=}k,\,S_n{=}s) \notag\\
  &\quad=P(K_n{=}k)\,P(W^n{=}k{-}s) \notag\\
  &\quad=|\calC_{n}|^{-1}q^{-n}
  =P(K_n{=}k)\,P(S_n{=}s), \notag
\end{align}
so $K_n\perp S_n$.
Since $I(W;Z)=0$, the pair $(K_n,W^n)$ is independent of $Z^n$, and hence
so is $(K_n,S_n)$.
Consequently, $I(K_n;S_n,Z^n)=0$ for every $n$.

\emph{Key rate.}
Since $\dim(\calC_{n})\ge n-m_n$, we have
\begin{equation}
  H(K_n)=\log|\calC_{n}|\ge(n-m_n)\log q,
\end{equation}
giving
$\liminf_{n\to\infty}\frac{1}{n}H(K_n)\ge(1-\rho)\log q>R_{\mathrm{key}}$.
Since $R_{\mathrm{key}}<R^{\star}_{\mathrm{univ}}$ was arbitrary, every
rate below $R^{\star}_{\mathrm{univ}}$ is universally achievable.
\end{IEEEproof}

\begin{remark}\label{rem:domain-limit}
The class $\mathcal{D}_{p_e}$ assumes \ac{iid} sources with uniform $W$ and zero leakage; the \ac{vpq} regime satisfies these only approximately. Non-uniformity of $W$ adds at most $\log q-\frac{1}{n}H(W^n)$ bit per symbol to the sketch leakage ($I(K_n;S_n)=H(S_n)-H(W^n)$), below $0.03$~bit at the entropies of Tables~\ref{tab:s4pls-syn} and~\ref{tab:s4pls-meas}; nonzero leakage is handled by Proposition~\ref{prop:bounded-leak}.
\end{remark}

\begin{prop}\label{prop:bounded-leak}
Assume the same setting as Theorem~\ref{thm:universal-rate}, except that the zero-leakage condition is relaxed to $I(W;Z)\le\epsilon$ for some $\epsilon\ge 0$.
Then the same code-offset construction achieves:
(i) reliability: $\Pr\{\hat{K}_n\neq K_n\}\to 0$;
(ii) rate-$\epsilon$ leakage: $\frac{1}{n}I(K_n;S_n,Z^n)\le\epsilon$ for every $n$ (which reduces to weak secrecy as $\epsilon\to 0$);
(iii) key rate: for every $\delta>0$, the construction at target key rate $R^{\star}_{\mathrm{univ}}-\delta$ attains $\liminf_{n\to\infty}\frac{1}{n}H(K_n)\ge(1-h_q(p_e))\log q -\delta$.%

\end{prop}

\begin{IEEEproof}
Reliability follows identically from the achievability argument in
Theorem~\ref{thm:universal-rate}, since the code design and the
Hamming-ball decoder depend only on $p_e$ and the uniformity of $W$, not
on the value of $I(W;Z)$.
The proof that $K_n\perp S_n$ likewise uses only the uniformity of $W^n$
over $\GF(q)^n$ and the independence of $K_n$ from $W^n$, both of which
hold regardless of $I(W;Z)$. Hence $I(K_n;S_n)=0$ for every $n$.

It remains to bound the leakage through $Z^n$.
Since $I(K_n;S_n)=0$, we have
\begin{align}
  I(K_n;S_n,Z^n)
  &= I(K_n;Z^n\mid S_n)
   = I(W^n;Z^n\mid S_n) \notag\\
  &\le I(W^n;Z^n)
   = nI(W;Z)
   \le n\epsilon, \notag
\end{align}
where the second equality uses the bijection $K_n\mapsto W^n=K_n-S_n$
given $S_n$, the inequality follows from the Markov chain $S_n$--$W^n$--$Z^n$ (since $K_n\perp(W^n,Z^n)$ gives $S_n\perp Z^n\mid W^n$), and the last equality uses the i.i.d.\ assumption.
Dividing by $n$ gives $\frac{1}{n}I(K_n;S_n,Z^n)\le\epsilon$.
The key-rate statement is identical to Theorem~\ref{thm:universal-rate}.
\end{IEEEproof}

The per-symbol leakage bound of Proposition~\ref{prop:bounded-leak} shrinks as the \ac{vpq} training reduces $I(W;Z)$, and the zero-leakage case recovers the per-block perfect secrecy of Theorem~\ref{thm:universal-rate}.

\begin{remark}
    Unlike the conventional usage of secure sketch in \ac{plk} generation\cite{wang2011fast,zhang2016key}, where Bob reconstructs $W^n$ and both parties adopt a further privacy-amplification step\cite{bennett1995generalized} to remove the information leaked by the public message, our proposed key reconciliation uses the randomly sampled codeword as the final secret key without any additional steps. The security is guaranteed directly by the \ac{vpq} objective: Theorem~\ref{thm:universal-rate} shows that, under zero leakage, the code-offset sketch achieves perfect secrecy at the optimal universal rate, and Proposition~\ref{prop:bounded-leak} bounds the key's per-symbol leakage by $I(W;Z)$ when the leakage is nonzero.%

\end{remark}

\subsection{Case Study: PLK Generation from Fading Channels}

As a canonical \ac{plk} generation scenario, we study the case of fading channels, where the estimated wireless channels at Alice and Bob are modeled as two correlated Gaussian random variables:
\begin{align}
    X = H+W_{1}, \quad Y = H+W_{2},
\end{align}
where $H\sim \calN(0, P)$ is the true channel between Alice and Bob, $W_{1}\sim \calN(0, N_{1})$ and $W_{2}\sim \calN(0, N_{2})$ are independent \ac{awgn}, and each vector $X, Y \in \RR^d$ consists of $d=8$ \ac{iid} such scalar samples. We consider three Eve cases (absent, uncorrelated, and correlated) corresponding to $Z=\varnothing$, independent Gaussian noise, or $Z = H + W_{3}$ with $W_{3}\sim \calN(0, N_{3})$; we set $P=N_3=0$~\si{dBm}.
To evaluate \ac{snr} robustness in a single sweep, we adopt the mixed-\ac{snr} training methodology of~\cite{guo2025physical}: the per-sample legitimate noise variance is drawn uniformly, $N_1 = N_2 \sim \mathrm{Uniform}[-30, 0]$~\si{dBm}, so that $\mathrm{SNR}_{\mathrm{ab}} := P/N_1$ covers $[0, 30]$~\si{dB} at every training step, and the trained encoder is tested at specific \ac{snr} points. Algorithm~\ref{alg:vske} is first applied to learn the encoders at both parties, and the Reed--Solomon codes are then used to reconcile the outputs into the final \acp{plk}. We compare the proposed \ac{vpq} at $|\calW|\in\{16,32,64,128,256\}$ against three reference methods: conventional \ac{plk} generation (Conv.\ KG)~\cite{zhang2016key} based on \ac{cdf} quantization and Gray coding; the mutual-information-driven autoencoder (MIAE)~\cite{guo2025physical}; and the domain-adversarial autoencoder (DAAE)~\cite{zhou2023daae}. MIAE and DAAE output continuous eight-dimensional features, quantized after training by the per-dimension \ac{cdf} rule at one bit per dimension, so both operate at $|\calW|=2^{8}=256$. The \ac{vpq} encoder $p_\theta$ is an $8$-layer \ac{fcn} with $1024$ units per layer, batch normalization, and ReLU activation; the adversarial predictor $p_\psi$ is an $8$-layer \ac{fcn} with $2048$ units per layer; batch size is $B=2048$ and the \ac{ema} factor is $\alpha=0.6$ without Eve and $\alpha=0.9$ with Eve. The mismatch weight is $\lambda_1=1$ in the no-Eve and uncorrelated-Eve scenarios; for the correlated Eve, $\lambda_1=2$ when $|\calW|\le 64$ and $\lambda_1=4$ when $|\calW|\ge 128$, while $\lambda_2$ follows the adaptive rule~\eqref{eq:lambda-update}. Optimization uses Adam with learning rate $3\times 10^{-5}$; the source is synthesized on the fly at $409\,600$ samples per epoch, the no-Eve and uncorrelated-Eve encoders are trained for $300$ epochs, and the correlated-Eve encoder is pretrained without the adversary for $50$ epochs and then trained adversarially up to epoch $1000$. Reconciliation uses Reed--Solomon codes over $\mathrm{GF}(|\calW|)$ with blocklength $n=|\calW|-1$ and message length $m$ swept.

\begin{figure*}[t]
    \centering
    \includegraphics[width=0.95\linewidth]{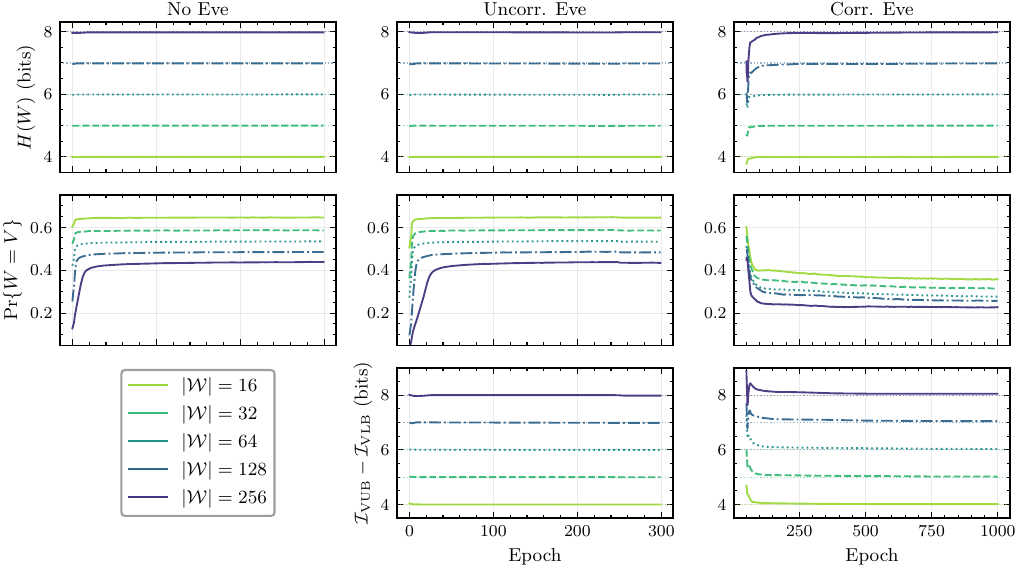}
    \caption{Training dynamics of the \ac{vpq} encoder on fading channels: encoder entropy $H(W)$ (top), agreement rate $\Pr\{W=V\}$ (middle), and observable leakage gap $\calI_{\mathrm{VUB}}-\calI_{\mathrm{VLB}}$ (bottom) over epochs, for the three Eve scenarios. Horizontal dotted lines mark $\log|\calW|$, which by Remark~\ref{rem:lautum} coincides with the zero-leakage signature of the gap.}
    \label{fig:fading_conv}
\end{figure*}

The \ac{vpq} encoders and the predictor are updated alternately as in Algorithm~\ref{alg:vske}. Fig.~\ref{fig:fading_conv} shows the training dynamics: the encoder entropy $H(W)$ converges to near $\log|\calW|$ in all scenarios within tens of epochs, and the agreement rate $\Pr\{W=V\}$ stabilizes quickly after the pretrain-to-adversarial transition. The observable gap $\calI_{\mathrm{VUB}} - \calI_{\mathrm{VLB}}$ settles near $H(W) = \log|\calW|$ for every alphabet, which, by Remark~\ref{rem:lautum}, is the empirical signature of the zero-leakage regime.

\begin{figure}[t]
    \centering
    \includegraphics[width=\columnwidth]{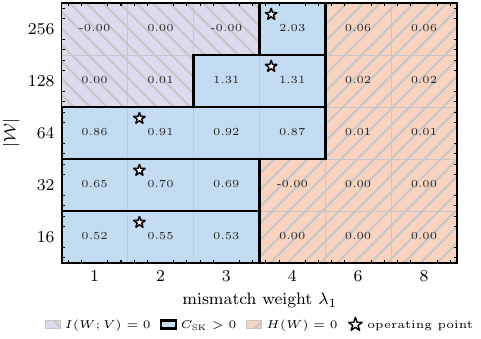}
    \caption{Stability of the \ac{vpq} training under the correlated Eve,
    one run of $1000$ epochs per cell, labeled by the secret-key rate
    $C_{\mathrm{SK}}$ in bits per symbol it ends at. Stars mark the $\lambda_1$
    used at each alphabet in this paper.}
    \label{fig:stability}
\end{figure}

\emph{Stability region.} Fig.~\ref{fig:stability} sweeps
$\lambda_1\in\{1,2,3,4,6,8\}$ against $|\calW|\in\{16,32,64,128,256\}$ under the
correlated-Eve schedule, $30$ runs of $1000$ epochs each. Each run's label is
its final $C_{\mathrm{SK}}=I(W;V)-\calI_{\mathrm{VUB}}$ over the
mixed-\ac{snr} training stream, read from the co-trained predictor, and sits
below the fixed-\ac{snr} values of Table~\ref{tab:maxleak}. A run is
usable when its label exceeds $0.1$~bit (every run ends below $0.07$ or above
$0.52$), and the failed runs are classified by $H(W)$. Each
alphabet is usable on an interval of $\lambda_1$ and fails on either side of it:
above the interval the encoders put all their mass on one symbol, below it $W$
and $V$ are independent, and either way $I(W;V)=0$.
The lower edge moves upward with $|\calW|$: the mismatch loss
is bounded in $[-1,0]$ for every alphabet while the uniformity loss spans
$[-\log|\calW|/(1-\alpha),0]$, so a larger alphabet needs a larger $\lambda_1$
before the two terms are comparable. The values used here, $\lambda_1=2$ for
$|\calW|\le64$ and $\lambda_1=4$ for $|\calW|\ge128$, lie inside the interval at
every alphabet; at $|\calW|=256$ the interval is one grid step wide.%

Too large a $\lambda_2$ fails as well: multiplying it by four drops
$C_{\mathrm{SK}}$ to $0.02$~bit at $|\calW|=64$ and $0.04$ at $|\calW|=256$.
Within the usable intervals of Fig.~\ref{fig:stability} the runs are flat over the second half of the adversarial training: the
standard deviation of $H(W)$ is at most $0.008$~bit and that of
$\Pr\{W{=}V\}$ at most $0.006$.%

After training, we evaluate on test \acp{snr} $\mathrm{SNR}_{\mathrm{ab}}\in\{0,5,\ldots,30\}$~\si{dB}, drawing $10^{6}$ fresh source samples at each \ac{snr}; in the Eve scenarios we additionally retrain a fresh $p_\psi$ on the frozen encoder outputs to obtain a tight estimate of $I(W;Z)$ and the induced-source secret-key rate $C_{\mathrm{SK}} = I(W;V)-I(W;Z)$.  This estimate is a variational lower bound, so it understates leakage if anything; the maximal leakage reported in Table~\ref{tab:maxleak} bounds $I(W;Z)$ from above.

\begin{figure}[t]
    \centering
    \includegraphics[width=\columnwidth]{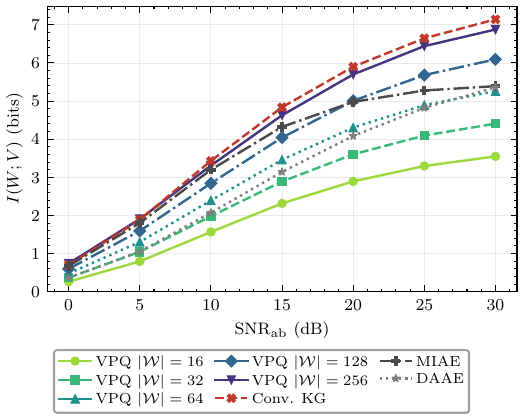}
    \caption{No-Eve case: mutual information $I(W;V)$ versus $\mathrm{SNR}_{\mathrm{ab}}$ for the three reference methods and \ac{vpq} at $|\calW|\in\{16,32,64,128,256\}$.}
    \label{fig:fading_snr_no_eve}
\end{figure}

\emph{No-Eve case} (Fig.~\ref{fig:fading_snr_no_eve}). $I(W;V)$ rises monotonically with \ac{snr} for every method, and, since $I(W;Z)=0$ holds automatically, Theorem~\ref{thm:vpq-keyrate} identifies $I(W;V)$ with the induced-source one-way secret-key capacity. \ac{vpq} with $|\calW|=256$ tracks the Conv.\ KG benchmark within $0.25$~bits at $\mathrm{SNR}_{\mathrm{ab}}\ge 15$~\si{dB}, while the feature-based autoencoders MIAE and DAAE plateau below \ac{vpq} $|\calW|=64$.

\begin{figure}[t]
    \centering
    \includegraphics[width=\columnwidth]{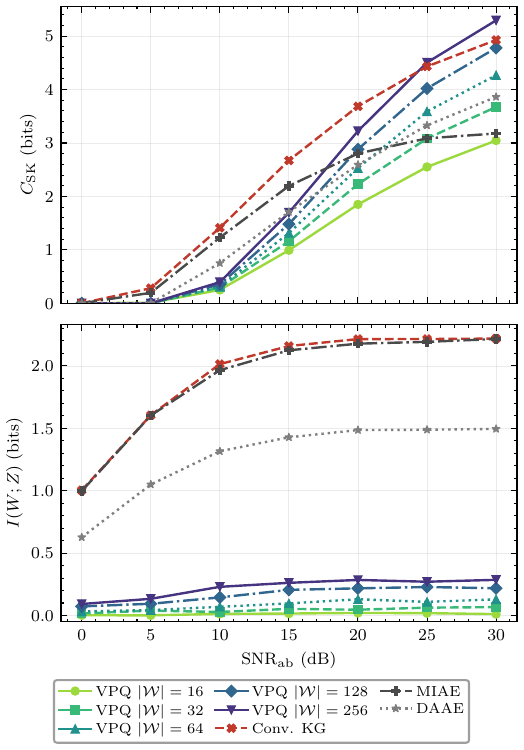}
    \caption{Correlated-Eve case: secret-key rate $C_{\mathrm{SK}}=I(W;V)-I(W;Z)$ (top) and Eve leakage $I(W;Z)$ (bottom) versus $\mathrm{SNR}_{\mathrm{ab}}$. \ac{vpq} keeps $I(W;Z)$ below $0.3$~bits for every alphabet; the baselines leak up to $2.2$~bits.}
    \label{fig:fading_snr_corr_eve}
\end{figure}

\emph{Correlated-Eve case} (Fig.~\ref{fig:fading_snr_corr_eve}). Every \ac{vpq} alphabet keeps $I(W;Z)\le 0.3$~bits uniformly over \ac{snr}, so the code-offset construction of Proposition~\ref{prop:bounded-leak} delivers keys without privacy amplification, with per-symbol key leakage at most $I(W;Z)$. The three baselines all leak $1.5$--$2.2$~bits per symbol at $20$~\si{dB} and above.
The comparison separates into two
effects. The fading source of this section is
a collection of independent scalar Gaussians, for which per-dimension \ac{cdf}
quantization is close to optimal; under the correlated Eve, Conv.\ KG extracts
more shared information than \ac{vpq}. Its quantizer is, however, fixed before
any eavesdropper is considered, so the correlation between the observation and
Eve's side information survives quantization, whereas the \ac{vpq} encoder is
trained against a predictor at Eve and leaks at least five times less at every
\ac{snr}. The
two effects work against each other in $C_{\mathrm{SK}}=I(W;V)-I(W;Z)$: the
shared information wins up to $\mathrm{SNR}_{\mathrm{ab}}=20$~\si{dB}, above
which \ac{vpq} overtakes Conv.\ KG (Fig.~\ref{fig:fading_snr_corr_eve}); at
$20$~\si{dB} \ac{vpq} also leaks least under the worst-case measure
(Table~\ref{tab:maxleak}). The advantage
Conv.\ KG holds here is tied to this source: its \ac{cdf}-based quantizer is
defined only on scalar observables with tractable marginal \ac{cdf}s, and has no
direct instantiation on the paired \ac{ra} maps of Section~IV.

\emph{Post-hoc quantization.} Whether a better quantizer would close the gap
of MIAE and DAAE to \ac{vpq} can be tested
directly: we freeze the trained encoders, change only the quantizer, and
evaluate every variant at $|\calW|=256$ and
$\mathrm{SNR}_{\mathrm{ab}}=20$~\si{dB} under the correlated Eve, on $10^{6}$
fresh source samples, with the predictor retrained on the variant's symbols.
Two variants are tested. The first replaces the per-dimension \ac{cdf} rule by
$k$-means over the eight feature dimensions jointly, at the same alphabet
size. The second keeps the \ac{cdf} rule but adds a guard band of width
$\beta$ around every quantization threshold and discards the samples that fall
inside it\cite{adil2021quantization}; $\beta$ is measured in per-dimension
probability mass.

Table~\ref{tab:ablation} shows that neither variant closes the gap. Joint
$k$-means changes $C_{\mathrm{SK}}$ by $-0.36$~bit/symbol for MIAE and by
$+0.06$ for DAAE, which still leaves DAAE $0.58$~bit below \ac{vpq}. At this
alphabet the $10^{6}$-sample joint histogram carries a plug-in bias near
$0.05$~bit, so the $k$-means deltas are read at that scale. The guard
band lowers the mismatch rate as intended: MIAE's $p_e$ falls from $0.53$ to
$0.22$ at $\beta=20\%$ and to $0.10$ at $\beta=40\%$, and its
$C_{\mathrm{SK}}$ rises from $2.80$ to $3.72$~bit per surviving symbol. A
sample survives only if it avoids the band in all eight dimensions: $16.0\%$
of the observations remain at $\beta=20\%$ and $1.3\%$ at $\beta=40\%$.
Averaged over all observations, the guard-banded MIAE delivers $0.60$ and
$0.03$~bit, far below the $3.224$~bit/symbol of \ac{vpq}. The $\beta=40\%$
rows rest on $13\,000$ and $206\,000$ surviving samples against $65\,536$
joint-histogram bins, which biases their $I(W;V)$ upward, in the guard band's
favor.

\begin{table}[t]
\caption{Post-hoc quantizer ablation on the frozen MIAE and DAAE encoders, at
$|\calW|=256$ and $\mathrm{SNR}_{\mathrm{ab}}=20$~\si{dB} under the correlated
Eve. The last column weights $C_{\mathrm{SK}}$ by the fraction of observations
the quantizer retains.}
\label{tab:ablation}
\centering
\footnotesize
\setlength{\tabcolsep}{3pt}
\renewcommand{\arraystretch}{1.2}
\begin{tabular}{llccccc}
\hline
& Quantizer & $I(W;V)$ & $p_e$ & $I(W;Z)$ & kept & $C_{\mathrm{SK}}$/obs. \\
\hline\hline
\multirow{4}{*}{MIAE}
 & \ac{cdf}, 1 bit/dim & $4.981$ & $0.529$ & $2.178$ & $1.000$ & $2.803$ \\
 & $k$-means, joint    & $4.799$ & $0.456$ & $2.351$ & $1.000$ & $2.448$ \\
 & guard band $20\%$   & $6.757$ & $0.221$ & $3.040$ & $0.160$ & $0.595$ \\
 & guard band $40\%$   & $7.298$ & $0.096$ & $4.580$ & $0.013$ & $0.034$ \\
\hline
\multirow{4}{*}{DAAE}
 & \ac{cdf}, 1 bit/dim & $4.078$ & $0.360$ & $1.486$ & $1.000$ & $2.592$ \\
 & $k$-means, joint    & $3.955$ & $0.528$ & $1.306$ & $1.000$ & $2.649$ \\
 & guard band $20\%$   & $2.927$ & $0.143$ & $1.215$ & $0.321$ & $0.549$ \\
 & guard band $40\%$   & $1.886$ & $0.068$ & $0.819$ & $0.206$ & $0.220$ \\
\hline
\end{tabular}
\end{table}

\begin{figure*}[t]
    \centering
    \includegraphics[width=0.95\linewidth]{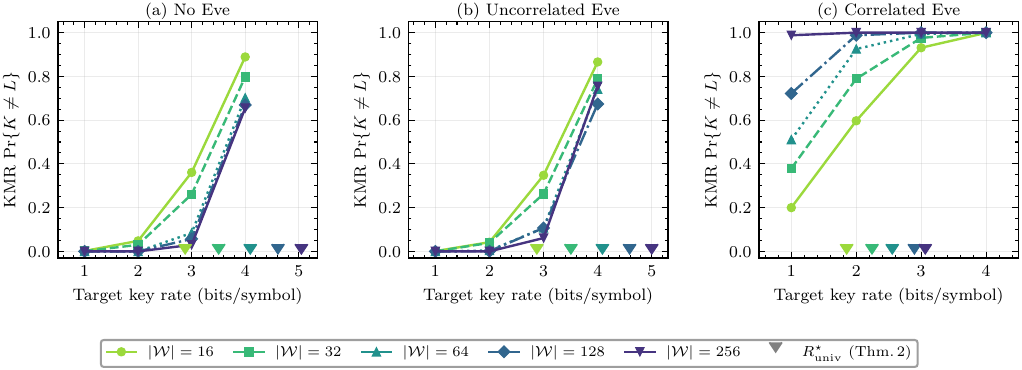}
    \caption{Key mismatch rate (KMR) versus target key rate for the three Eve scenarios at $\mathrm{SNR}_{\mathrm{ab}}=20$~\si{dB}, for all \ac{vpq} alphabets $|\calW|\in\{16,32,64,128,256\}$. Triangle markers on each $x$-axis indicate the per-alphabet universal benchmark $R^{\star}_{\mathrm{univ}}=(1-h_q(p_e))\log q$ from Theorem~\ref{thm:universal-rate}, computed from the measured mismatch rate $p_e=1-\Pr\{W=V\}$.}
    \label{fig:fading_kmr}
\end{figure*}

\emph{Cost of privacy amplification.} The baselines need a
privacy-amplification stage that \ac{vpq} does not, since
Proposition~\ref{prop:bounded-leak} bounds the leakage of its reconciled key by
$I(W;Z)$ at every blocklength. A method with higher leakage hashes $N$ of its
symbols with a function drawn at random from a universal$_2$
family\cite{bennett1995generalized} and announced publicly. By
the leftover hash lemma\cite{renner2005simple} an $\ell$-bit output is within statistical
distance $\varepsilon$ of uniform given $Z^{N}$ provided
$\ell \le H_{\min}^{\varepsilon}(W^{N}|Z^{N}) - 2\log_2(1/\varepsilon)$, where
$H_{\min}^{\varepsilon}$ is the $\varepsilon$-smooth conditional min-entropy.
For an \ac{iid} source, $-\log_2 p(W^{N}|Z^{N})$ has mean $N H(W|Z)$ and
standard deviation $\sqrt{N \sigma^2_{W|Z}}$ with
$\sigma^2_{W|Z} = \mathrm{Var}[-\log_2 p(W|Z)]$ the conditional varentropy,
so\cite{tomamichel2013hierarchy}
\begin{equation}\label{eq:hmin-expansion}
\begin{split}
  &H_{\min}^{\varepsilon}(W^{N}|Z^{N}) = N H(W|Z) \\
  &\qquad - \sqrt{N \sigma^2_{W|Z}}\,\Phi^{-1}(1-\varepsilon) + O(\log N),
\end{split}
\end{equation}
with $\Phi^{-1}$ the inverse standard normal \ac{cdf}. We measure
$\sigma^2_{W|Z}$ from the predictor $p_\psi$. Per symbol the extractor therefore
loses
\begin{equation}\label{eq:pa-deficit}
  \Delta(N) = \sqrt{\frac{\sigma^2_{W|Z}}{N}}\,\Phi^{-1}(1-\varepsilon)
            + \frac{2\log_2(1/\varepsilon)}{N},
\end{equation}
an approximation rather than a bound, since the $O(\log N)$ term of
\eqref{eq:hmin-expansion} is dropped; a baseline delivers
$C_{\mathrm{SK}} - \Delta(N)$ where \ac{vpq} delivers $C_{\mathrm{SK}}$.
$\Delta(N)$ is a finite-blocklength cost on top of the leakage already
subtracted inside $C_{\mathrm{SK}}$ (Theorem~\ref{thm:vpq-keyrate}), and it
vanishes as $N$ grows.
The hashed key meets a statistical-distance criterion while
Proposition~\ref{prop:bounded-leak} bounds a per-symbol mutual information;
the distinction has no asymptotic consequence, since strong secrecy is
achievable at the weak-secrecy rate\cite{maurer2000information}.

Table~\ref{tab:pa-deficit} evaluates $\Delta(N)$ at $\varepsilon=10^{-9}$, and
over $\varepsilon\in[10^{-12},10^{-6}]$ it varies by less than a factor of two.
Conv.\ KG holds the higher asymptotic rate, $3.687$ against $3.224$~bit/symbol
for \ac{vpq}. The two rates meet near $N\approx9\times10^{2}$~symbols, where $\Delta(N)$
equals their gap of $0.463$~bit/symbol; below that blocklength \ac{vpq}
delivers more. MIAE and DAAE stay below \ac{vpq} at every $N$; their
asymptotic rates are $2.803$ and $2.592$~bit/symbol.

\begin{table}[t]
\caption{Privacy-amplification deficit $\Delta(N)$ of~\eqref{eq:pa-deficit},
in bit/symbol, versus the extractor block length $N$. Correlated Eve,
$|\calW|=256$, $\mathrm{SNR}_{\mathrm{ab}}=20$~\si{dB}, $\varepsilon=10^{-9}$.}
\label{tab:pa-deficit}
\centering
\setlength{\tabcolsep}{6pt}
\renewcommand{\arraystretch}{1.2}
\begin{tabular}{lccccc}
\hline
Method & $\sigma^2_{W|Z}$ & $N{=}10^{3}$ & $10^{4}$ & $10^{5}$ & $10^{6}$ \\
\hline\hline
Conv.\ KG & $4.116$ & $0.445$ & $0.128$ & $0.039$ & $0.012$ \\
MIAE      & $4.125$ & $0.445$ & $0.128$ & $0.039$ & $0.012$ \\
DAAE      & $5.757$ & $0.515$ & $0.150$ & $0.046$ & $0.014$ \\
\hline
\end{tabular}
\end{table}

\emph{Maximal leakage.} Besides $I(W;Z)$ we report a worst-case measure of
secrecy, the maximal leakage of~\cite{issa2020operational},
\begin{equation}\label{eq:maxleak}
  \calL_{\max}(W\!\to\!Z) = \log \sum_{z} \max_{w} p(z\mid w),
\end{equation}
the largest factor by which observing $Z$ multiplies the probability of
correctly guessing any (possibly randomized) function of $W$, on a logarithmic
scale. By
Bayes' rule $\calL_{\max}=\log\expcs{p(z)}{\max_{w} p(w\mid z)/p(w)}$, which we
estimate from the predictor $p_\psi$ that already gives $\calI_{\mathrm{VLB}}$
and from the empirical symbol frequencies over the $N_{\mathrm{t}}$ test samples,
\begin{equation}\label{eq:maxleak-est}
  \hat{\calL}_{\max}
  = \log \frac{1}{N_{\mathrm{t}}} \sum_{i=1}^{N_{\mathrm{t}}}
    \max_{w\,:\,p(w)>0} \frac{p_\psi(w\mid z_i)}{p(w)} .
\end{equation}
Table~\ref{tab:maxleak} evaluates it at
$\mathrm{SNR}_{\mathrm{ab}}=20$~\si{dB} under the correlated Eve. A maximum over a finite
sample is positive even under independence: with an uncorrelated Eve the same
estimator returns $0.06$--$0.52$ at $I(W;Z)=0.000$, and every entry of the table
is at least $2.7$ times the floor of the same encoder. \ac{vpq} stays below $I(W;V)$ at every alphabet,
while the three baselines reach $4.7$--$4.8$; for MIAE that is within
$0.3$~bit of its own $I(W;V)$, for DAAE above it.

\begin{table}[t]
\caption{Maximal leakage~\eqref{eq:maxleak} at
$\mathrm{SNR}_{\mathrm{ab}}=20$~\si{dB} under the correlated Eve, in bits.}
\label{tab:maxleak}
\centering
\setlength{\tabcolsep}{6pt}
\renewcommand{\arraystretch}{1.2}
\begin{tabular}{lcccc}
\hline
Method & $|\calW|$ & $I(W;V)$ & $I(W;Z)$ & $\calL_{\max}$ \\
\hline\hline
\ac{vpq} & $16$  & $1.870$ & $0.020$ & $0.512$ \\
\ac{vpq} & $32$  & $2.275$ & $0.046$ & $0.867$ \\
\ac{vpq} & $64$  & $2.660$ & $0.130$ & $1.380$ \\
\ac{vpq} & $128$ & $3.100$ & $0.217$ & $1.997$ \\
\ac{vpq} & $256$ & $3.509$ & $0.285$ & $2.451$ \\
Conv.\ KG & $256$ & $5.902$ & $2.215$ & $4.712$ \\
MIAE      & $256$ & $4.981$ & $2.178$ & $4.697$ \\
DAAE      & $256$ & $4.078$ & $1.486$ & $4.792$ \\
\hline
\end{tabular}
\end{table}

\emph{Reed--Solomon reconciliation} (Fig.~\ref{fig:fading_kmr}). We reconcile the trained encoder outputs at $\mathrm{SNR}_{\mathrm{ab}}=20$~\si{dB} with Reed--Solomon codes $\RS(n,m)$ of blocklength $n=|\calW|-1$, sweeping the message length $m$ to trace the \ac{kmr} against the target key rate $R=(m/n)\log|\calW|$. Triangle markers on each $x$-axis indicate the per-alphabet asymptotic bound $R^\star_{\mathrm{univ}}=(1-h_q(p_e))\log q$ from Theorem~\ref{thm:universal-rate}, evaluated at the measured mismatch rate $p_e=1-\Pr\{W=V\}$; by Theorem~\ref{thm:universal-rate}, no reconciliation scheme can be uniformly reliable and secure over $\mathcal{D}_{p_e}$ at rates above the triangle. The \ac{rs} cliffs sit strictly to the left of the corresponding triangles, which is the expected finite-blocklength gap between a length-$n$ linear code and the asymptotic universal limit. When $p_e$ is moderate (no-Eve and uncorrelated-Eve panels) the ordering across alphabets is consistent: larger $|\calW|$ pushes both $R^\star_{\mathrm{univ}}$ and the cliff to the right, so higher target rates become reachable by increasing the alphabet. In the correlated-Eve panel, however, tightening the leakage constraint inflates $p_e$ more severely for large $|\calW|$, which collapses $R^\star_{\mathrm{univ}}$ (for $|\calW|=256$ even $R=1$ already yields $\mathrm{KMR}\approx 1$), so the ordering inverts and the smallest alphabets, with the lowest $p_e$ and therefore the largest $R^\star_{\mathrm{univ}}$, deliver the best low-rate operating point.

\emph{Complexity.} Per training iteration, the loss evaluation is dominated by the empirical \ac{vub} estimator~\eqref{eq:vub}, whose cross-batch double sum requires $\mathcal{O}(B^2|\calW|)$ operations for a batch size $B$ and alphabet $|\calW|$, against $\mathcal{O}(B|\calW|)$ for the entropy, mismatch, and \ac{vlb} losses. The encoder and predictor forward/backward passes add $\mathcal{O}(B\, d_{\mathrm{FCN}}^{2})$ per hidden layer of width $d_{\mathrm{FCN}}$. Inference uses a single forward pass per sample at $\mathcal{O}(d_{\mathrm{FCN}}^{2})$ per layer.

Overall, these results demonstrate that the proposed framework can extract high-entropy keys from fading channels, with the leakage of the reconciled key bounded at the quantization stage (cf.\ Theorem~\ref{thm:universal-rate} and Proposition~\ref{prop:bounded-leak}), even in the presence of a correlated Eve.

%% file: paper-sections/s4pls.tex
Traditional \ac{plk} generation based on channel reciprocity requires two-way probing between Alice and Bob, which becomes expensive under high mobility where the channel coherence time is too short to accommodate the protocol. We instead propose a sensing-based \ac{plk} scheme enabled by \ac{isac}\cite{liu2022integrated}: Alice transmits, both Alice and Bob sense the common propagation environment, and the paired \ac{ra} maps they estimate serve as the \ac{cr} source. When Bob's position changes rapidly between updates, successive \ac{ra} maps become effectively independent, providing the fresh randomness required for \ac{plk} generation. The scheme is validated end to end on synthesized \ac{ra} maps derived from ray-traced 5G~NR channels and at the quantization stage on \ac{sdr} measurements collected in a furnished lab room and an anechoic chamber.

\begin{figure*}[t]
    \centering
    \begin{subfigure}[t]{0.32\textwidth}
        \centering
        \includegraphics[width=\textwidth]{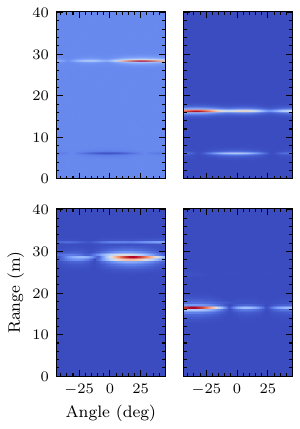}
        \caption{Synthesized from the DeepMIMO dataset.}
    \end{subfigure}\hfill
    \begin{subfigure}[t]{0.32\textwidth}
        \centering
        \includegraphics[width=\textwidth]{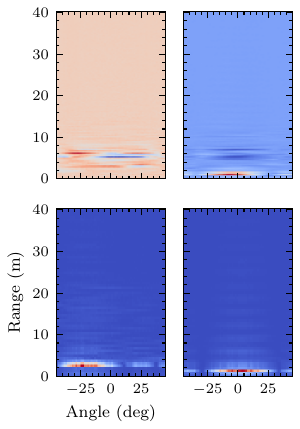}
        \caption{Measured in the lab room.}
    \end{subfigure}\hfill
    \begin{subfigure}[t]{0.32\textwidth}
        \centering
        \includegraphics[width=\textwidth]{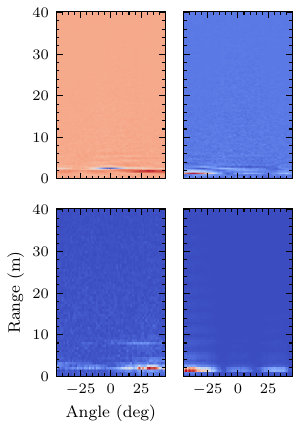}
        \caption{Measured in the anechoic chamber.}
    \end{subfigure}
    \caption{Examples of synthesized and measured \ac{ra} maps of Alice (above) and Bob (below).}
    \label{fig:ramap}
\end{figure*}

\subsection{Channel Model and \ac{ra} Map}

We consider an \ac{isac} system with a static transmitter Alice and a dynamic receiver Bob, both equipped with steerable beamformers; Alice additionally uses a co-located receiver beamformer to capture her own echo. For notational simplicity we describe 2D beamforming; the extension to 3D is straightforward. Alice maps data and pilots onto an \ac{ofdm} grid of $N_{\mathrm{sc}}$ subcarriers at spacing $\Delta f$ and transmits the time-domain signal $s(t)$ through a beam fixed at azimuth $\varphi$. Both parties receive through steerable beams at angle $\theta_\ast$, $\ast\in\{\ma,\mb\}$, yielding\cite{zhang2021overview}
\begin{equation}
    y_\ast(t) = \bm a_{\mr,\ast}^\herm(\theta_\ast)\sum_{l=0}^{L_\ast}\bm H_{\ast,l}(t)\,\bm a_\mt(\varphi)\, s(t-\tau_{\ast,l}) + n_\ast(t),
\end{equation}
where $\bm a_{\mathrm{Rx},\ast}$ and $\bm a_{\mathrm{Tx}}$ are the receive and transmit array steering vectors, and $\bm H_{\ast,l}$, $\tau_{\ast,l}$, and $\varphi_{\ast,l},\theta_{\ast,l}$ are the $l$-th path's gain matrix, delay, and \ac{aod}/\ac{aoa}. Denoting $l=0$ as the \ac{los} path and assuming Alice's and Bob's arrays are parallel, the echo path at Alice satisfies $\tau_{\ma,0}=2\tau_{\mb,0}$ and $\|\bm H_{\ma,0}\|\propto\sqrt{\rcs}\,\|\bm H_{\mb,0}\|/2$ through Bob's \ac{rcs}, and we require a \ac{cp} longer than the maximum path delay. Under clock synchronization and aligned demodulation, both parties obtain pilot-based least-squares \ac{ofdm} channel estimates $\hat{\bm H}_\ast(\theta_\ast)[\nsub,\nsym]$ at subcarrier index $\nsub$ and symbol index $\nsym$.

To form an \ac{ra} map we pick a fixed \ac{ofdm} symbol (canceling Doppler), let $\hat{\bm h}_\ast(\theta_\ast)[\nsub]\!=\!\hat{\bm H}_\ast(\theta_\ast)[\nsub,0]$, and apply an $\nifft$-point \ac{ifft} to the zero-padded estimate,
\begin{equation}
    \bm r_\ast(\theta_\ast)[n] = \frac{1}{\nifft}\left|\sum_{n'=0}^{\nifft-1} \hat{\bm h}_\ast(\theta_\ast)[n']\, e^{j2\pi\frac{nn'}{\nifft}}\right|^{2};
\end{equation}
a peak at index $\hat n_\ast$ corresponds to a scatterer at range
\begin{equation}
    \hat d_\mb = \frac{\hat n_\mb c_0}{\Delta f\,\nifft},\quad \hat d_\ma = \frac{\hat n_\ma c_0}{2\Delta f\,\nifft},
\end{equation}
with $c_0$ the speed of light and the factor $1/2$ at Alice accounting for the round-trip propagation. Sweeping $\theta_\ast$ over $\na$ beams and stacking the $\nr$ range bins yields the \ac{ra} map $\bX_\ast\in\RR^{\nr\times\na}$. When Alice subtracts a clutter-only \ac{ra} map measured without Bob, the strongest surviving peak corresponds to the \ac{los}-plus-echo component, so the peaks at Alice and Bob coincide in $(\hat d,\hat\theta)$ and the paired \ac{ra} maps are highly correlated (Fig.~\ref{fig:ramap}).

\begin{remark}
When Alice and Bob run on independent references, their clocks drift apart and the range axes of the paired \ac{ra} maps slide against each other. An initial reference transmission from Alice aligns the two frames; the misalignment then grows at the references' fractional frequency offset $\epsilon$, and the maps stop lining up once it reaches one range bin, $c_0/2B$ at occupied bandwidth $B$. Re-anchoring before that costs one acquisition, during which no key is produced. Taking half a range bin as the trigger, the anchor period is $T_{\mathrm{ref}} = 1/(2B\epsilon)$ and an acquisition of duration $T_{\mathrm{acq}}$ leaves
\begin{equation}\label{eq:sync}
  R_{\mathrm{eff}} = \Bigl(1 - \frac{T_{\mathrm{acq}}}{T_{\mathrm{acq}}+T_{\mathrm{ref}}}\Bigr) R_{\mathrm{key}}.
\end{equation}
One acquisition sweeps $64$ beams at one \ac{ofdm} symbol each, so $T_{\mathrm{acq}}=\SI{571}{\micro\second}$ for the front end used here. At a relative offset of $0.05$~ppm, the wide-area per-device limit of TS~38.104\cite{3gpp_ts_38104}, the system re-anchors every \SI{25}{ms} and loses $2.2\%$ of the rate. At $0.1$~ppm, which is both the medium/local-area limit and the worst offset between two compliant wide-area references, it loses $4.4\%$.

The acquisition time $T_{\mathrm{acq}}$ is the only term the front end sets. A fully digital array forms all $64$ beams from a single snapshot, $T_{\mathrm{acq}}=\SI{8.92}{\micro\second}$, and the same two clocks then cost $0.036\%$ and $0.071\%$. The overhead is set by how long one acquisition takes, not by the key-generation scheme. The angular frame slips with the same clocks through the beam schedule, one \ac{ofdm} symbol per beam. Its trigger, half a symbol, is three orders of magnitude looser than the range frame's $1/(2B)$, so the range frame binds.
\end{remark}

\subsection{Data Collection}
\label{sec:dataset}

\emph{Synthesized dataset.} Ray-traced path parameters of the DeepMIMO O1\_28 scenario\cite{alkhateeb2019deepmimo,Remcom} are used to synthesize the \ac{ra} maps in MATLAB with the 5G~NR Toolbox: base stations play Alice, users play Bob, and Alice's echo path is injected into her self-interference channel by reusing the Alice-to-Bob \ac{los} ray with doubled delay, $-\SI{6}{dB}$ gain, and a randomly sampled $\rcs$. The path parameters then drive a 3GPP \ac{cdl} channel model\cite{3gpp_tr_38901}; Alice's sectorized beamformers are $30^\circ$ down-tilt and Bob's $30^\circ$ up-tilt to ensure array parallelism. All remaining parameters (carrier, waveform, beam grid) are listed in Section~\ref{sec:experiments}. The scenario yields $102\,090$ paired maps; $80\%$ are used for training, $10\%$ for validation, and $10\%$ for test.

\begin{figure}[!tbp]
    \centering
    \begin{subfigure}[b]{\columnwidth}
        \centering
        \includegraphics[width=\columnwidth]{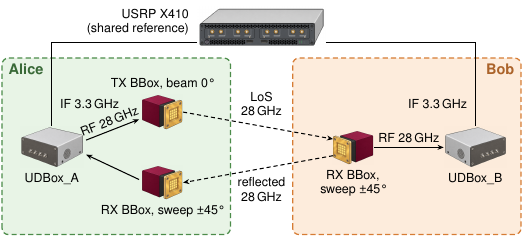}
        \caption{Signal-flow diagram.}
        \label{fig:hw_diag}
    \end{subfigure}
    \par\vspace{2mm}
    \begin{subfigure}[b]{0.48\columnwidth}
        \centering
        \includegraphics[width=\textwidth]{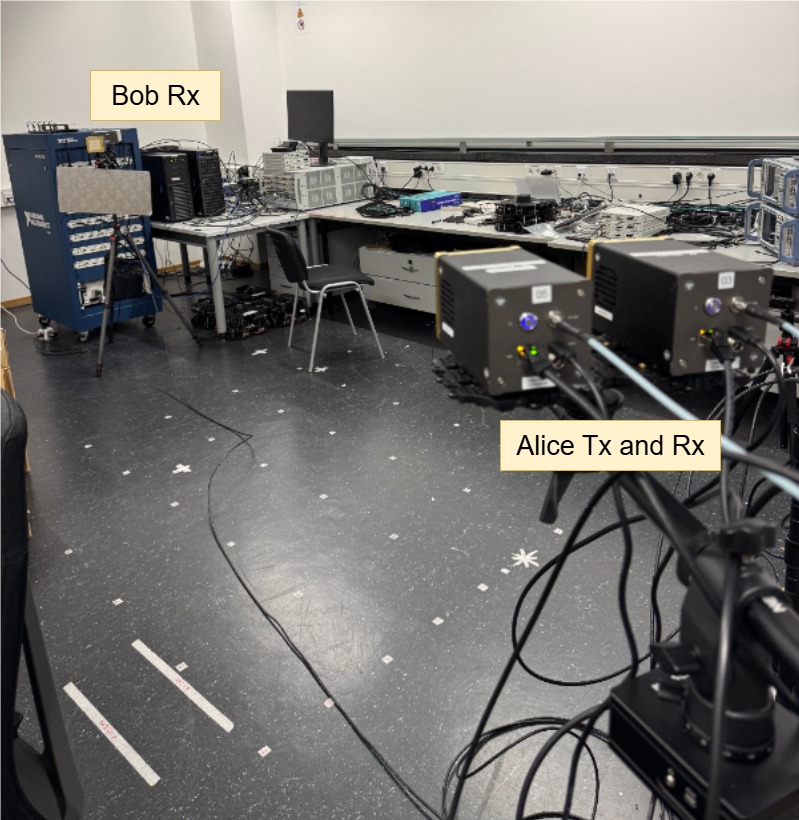}
        \caption{Lab room.}
        \label{fig:hw_lab}
    \end{subfigure}\hfill
    \begin{subfigure}[b]{0.48\columnwidth}
        \centering
        \includegraphics[width=\textwidth]{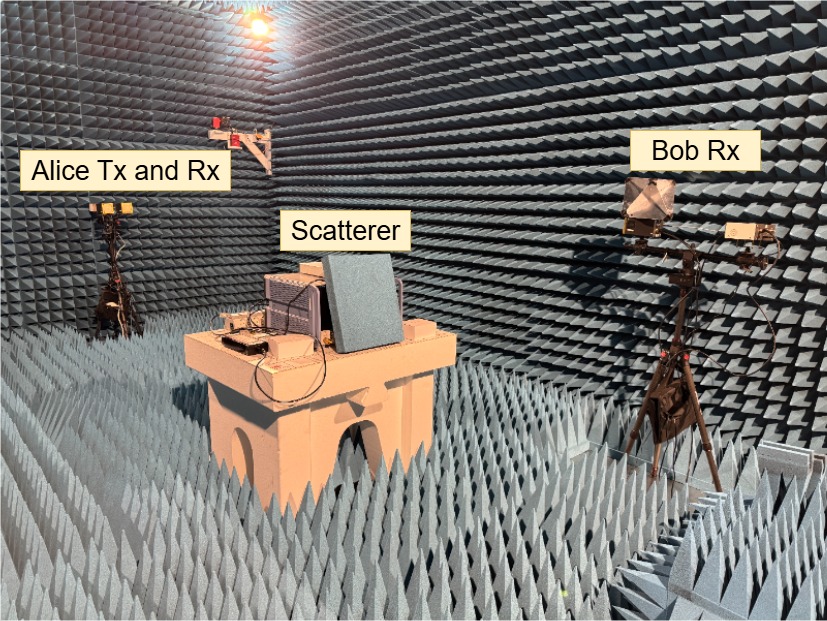}
        \caption{Anechoic chamber.}
        \label{fig:hw_chamber}
    \end{subfigure}
    \caption{Hardware setup for the \ac{ra} map measurement testbed: signal-flow diagram (a) and two physical environments (b, c).}
    \label{fig:hw}
\end{figure}

\emph{Real-world measurements.} We reproduce the synthesized-dataset geometry with an \ac{sdr} testbed at the \ac{aces} Lab of the \ac{tum}, operating in both a furnished lab room and an anechoic chamber. As shown in Fig.~\ref{fig:hw}(a), Alice uses two TMYTEK BBox mmWave beamformers (one transmitter, one receiver) each fed by a TMYTEK UDBox up/down-converter, while Bob uses one BBox plus one UDBox on the receive side; both UDBoxes connect to the same \ac{usrp} X410 so that transmission and reception are timing-synchronized. The baseband \ac{pdsch} signal is generated on the host PC, upconverted from \SI{3.3}{GHz} \ac{if} to \SI{28}{GHz} through the UDBox--BBox chain, and the two receivers sweep $64$ azimuth beams in $[-45^\circ,45^\circ]$ with the transmit beam fixed at $0^\circ$, so the measurement setup exactly mirrors the simulation. For each environment we record paired \ac{ra} maps with Bob placed at $41$ locations in the lab room and $43$ in the anechoic chamber, four repeated measurements at each, then a clutter-only scan with Bob removed to subtract environmental scattering. Two of the four measurements at each position are used for fine-tuning and the other two for test.

Fig.~\ref{fig:ramap} shows representative synthesized and measured \ac{ra} maps: the two datasets share the same peak geometry, while the lab room is noisier than the chamber due to extra scatterers. Both datasets confirm that the paired maps are correlated enough to serve as the \ac{plk} source.

\subsection{Experiments}
\label{sec:experiments}

\begin{table*}[!t]
\centering
\caption{Test results of \ac{vpq} models trained on the synthesized dataset. The correlated-Eve case spans the full $(\Delta D,\Delta\Theta)$ grid together with the perfect-position point $(0,0^\circ)$.}
\label{tab:s4pls-syn}
\footnotesize
\setlength{\tabcolsep}{9pt}
\renewcommand{\arraystretch}{1.25}
\begin{tabular}{l | c c | c c c c }
\hline
Case   & $\Delta D$ (meter) & $\Delta\Theta$ (degree) & $H(W)$ & $\pr{W=V}$ & $\calI_{\mathrm{VLB}}$ & $\calI_{\mathrm{VUB}}$\\
\hline\hline
No Eve              & -   & -   & $3.971$ & $0.953$ & -        & -        \\
\hline
Uncorrelated Eve    & -   & -   & $3.970$ & $0.958$ & $-3.975$ & $-0.000$ \\
\hline
Correlated Eve (perfect) & $0$ & $0$ & $3.992$ & $0.716$ & $-3.536$ & $0.650$ \\
\hline
\multirow{12}{*}{Correlated Eve}
 & \multirow{3}{*}{$10$} & $15$ & $3.978$ & $0.904$ & $-3.839$ & $0.298$ \\
 &                       & $10$ & $3.988$ & $0.930$ & $-3.853$ & $0.305$ \\
 &                       & $5$  & $3.987$ & $0.939$ & $-3.848$ & $0.330$ \\
\cline{2-7}
 & \multirow{3}{*}{$5$}  & $15$ & $3.990$ & $0.897$ & $-3.862$ & $0.273$ \\
 &                       & $10$ & $3.987$ & $0.917$ & $-3.846$ & $0.307$ \\
 &                       & $5$  & $3.994$ & $0.909$ & $-3.828$ & $0.358$ \\
\cline{2-7}
 & \multirow{3}{*}{$3$}  & $15$ & $3.993$ & $0.873$ & $-3.798$ & $0.378$ \\
 &                       & $10$ & $3.993$ & $0.878$ & $-3.784$ & $0.407$ \\
 &                       & $5$  & $3.985$ & $0.854$ & $-3.783$ & $0.400$ \\
\cline{2-7}
 & \multirow{3}{*}{$1$}  & $15$ & $3.992$ & $0.784$ & $-3.659$ & $0.537$ \\
 &                       & $10$ & $3.997$ & $0.796$ & $-3.675$ & $0.515$ \\
 &                       & $5$  & $3.996$ & $0.786$ & $-3.691$ & $0.505$ \\
\hline
\end{tabular}
\end{table*}

The synthesized dataset is built from DeepMIMO O1\_28\cite{alkhateeb2019deepmimo} at \SI{28}{GHz} (band n257), using a 5G NR \ac{pdsch} waveform with \SI{120}{kHz} subcarrier spacing and $275$ resource blocks (the full \SI{400}{MHz} band); Alice and Bob are both equipped with a $4{\times}4$ uniform rectangular array at $0.5\lambda$ spacing, and the receiver sweeps $64$ azimuth beams over $[-45^\circ,45^\circ]$ with the transmit beam fixed at $0^\circ$. The bistatic echo path follows the 3GPP \ac{cdl} model\cite{3gpp_tr_38901} with doubled \ac{los} delay and $-\SI{6}{dB}$ gain;
per-sample receive-noise levels are drawn across a range that keeps the operating \ac{snr} in a realistic regime for 28~GHz sensing. Because the \ac{ra} maps are naturally sequential along the range axis, both Alice and Bob use independent standard Transformer encoders (4~layers, 8~heads, $d_{\mathrm{model}}=256$, with multi-head self-attention, learnable positional encoding, residual connections, and layer normalization, followed by a final linear+softmax layer producing categorical outputs on $\calW$); other sequence architectures such as CNN or RNN are equally admissible. We fix $|\calW|=16$ throughout and train with AdamW at learning rate $10^{-4}$ and batch size $2048$ for $1000$ epochs; when a correlated Eve is present we set $\lambda_1=\lambda_2=1$ and reserve the last $50$ epochs for an Eve-only predictor update. For the real-world datasets we freeze the backbone and fine-tune only the output linear layer with AdamW at learning rate $3\times 10^{-4}$ on a CosineAnnealingLR schedule for $1000$ epochs without early stopping; the fine-tune loss uses $\lambda_1=1$, $\lambda_2=2$. Eve's position uncertainty is swept over $\Delta D\in\{1,3,5,10\}$~m paired with $\Delta\Theta\in\{5,10,15\}^\circ$, together with the perfect-position point $(0,0^\circ)$.

\begin{table*}[!t]
\centering
\caption{Test results of the \ac{vpq} stage on the real-world \ac{ra} map datasets after fine-tuning, with $95\%$ confidence intervals from the delta method over the test split.}
\label{tab:s4pls-meas}
\footnotesize
\setlength{\tabcolsep}{3pt}
\renewcommand{\arraystretch}{1.3}
\begin{tabular}{l l | c c c c c }
\hline
Dataset & Case & $H(W)$ & $\pr{W=V}$ & $\calI_{\mathrm{VLB}}$ & $\calI_{\mathrm{VUB}}$ & $\calL_{\max}$\\
\hline\hline
\multirow{4}{*}{Lab room}
 & No Eve & $3.921$ \tiny{$[3.817, 4.000]$} & $0.780$ \tiny{$[0.717, 0.843]$} & - & - & - \\
 & Uncorr.\ Eve & $3.931$ \tiny{$[3.840, 4.000]$} & $0.685$ \tiny{$[0.617, 0.754]$} & $-3.957$ \tiny{$[-4.049, -3.866]$} & $-0.018$ \tiny{$[-0.044, 0.008]$} & $0.283$ \tiny{$[0.252, 0.313]$} \\
 & Corr.\ Eve $(1\,\mathrm{m}, 5^\circ)$ & $3.993$ \tiny{$[3.982, 4.000]$} & $0.320$ \tiny{$[0.283, 0.357]$} & $-3.885$ \tiny{$[-3.926, -3.843]$} & $0.185$ \tiny{$[0.143, 0.227]$} & $0.926$ \tiny{$[0.878, 0.974]$} \\
 & Corr.\ Eve (perfect) & $3.997$ \tiny{$[3.991, 4.000]$} & $0.112$ \tiny{$[0.093, 0.130]$} & $-3.980$ \tiny{$[-3.995, -3.965]$} & $0.037$ \tiny{$[0.021, 0.053]$} & $0.469$ \tiny{$[0.424, 0.514]$} \\
\hline
\multirow{4}{*}{Anechoic chamber}
 & No Eve & $3.824$ \tiny{$[3.674, 3.974]$} & $0.823$ \tiny{$[0.761, 0.886]$} & - & - & - \\
 & Uncorr.\ Eve & $3.922$ \tiny{$[3.822, 4.000]$} & $0.834$ \tiny{$[0.794, 0.875]$} & $-3.941$ \tiny{$[-4.040, -3.842]$} & $-0.014$ \tiny{$[-0.033, 0.004]$} & $0.172$ \tiny{$[0.126, 0.218]$} \\
 & Corr.\ Eve $(1\,\mathrm{m}, 5^\circ)$ & $3.998$ \tiny{$[3.994, 4.000]$} & $0.270$ \tiny{$[0.251, 0.290]$} & $-3.874$ \tiny{$[-3.903, -3.845]$} & $0.188$ \tiny{$[0.156, 0.220]$} & $0.960$ \tiny{$[0.904, 1.016]$} \\
 & Corr.\ Eve (perfect) & $3.998$ \tiny{$[3.995, 4.000]$} & $0.186$ \tiny{$[0.168, 0.204]$} & $-3.928$ \tiny{$[-3.946, -3.910]$} & $0.110$ \tiny{$[0.091, 0.128]$} & $0.775$ \tiny{$[0.733, 0.816]$} \\
\hline
\end{tabular}
\end{table*}

For the correlated Eve, $Z$ is an estimate of Bob's position with controllable uncertainty:
\begin{equation}\label{eq:eve-position}
    \hat d_\me = d + \Delta d,\quad \hat\theta_\me = \theta + \Delta\theta,
\end{equation}
with $\Delta d,\Delta\theta$ uniform on $[-\Delta D/2,\Delta D/2]$ and $[-\Delta\Theta/2,\Delta\Theta/2]$; the uncorrelated-Eve baseline takes $(\hat d_\me,\hat\theta_\me)$ independent of $(d,\theta)$. The perfect-position point $(\Delta D,\Delta\Theta)=(0,0^\circ)$, at which Bob's range and angle are known exactly, is the strongest adversary of this form.

On the synthesized dataset (Table~\ref{tab:s4pls-syn}), the \ac{vpq} outputs are nearly uniform ($H(W)\to\log|\calW|=4$) and agree with $\Pr\{W=V\}>0.9$ when Eve is absent, uncorrelated, or localized no better than $\Delta D=10$~m; the agreement drops and $\calI_{\mathrm{VLB}}/\calI_{\mathrm{VUB}}$ grow as Eve's position uncertainty shrinks. Even under the perfect-position Eve, the encoder still delivers $\Pr\{W=V\}=0.72$ with a leakage ceiling $\calI_{\mathrm{VUB}}=0.65$, showing that the shared randomness exploited by \ac{vpq} is carried by the multipath and scattering structure of the \ac{ra} maps rather than by Bob's position alone. Applying the Reed--Solomon code-offset reconciliation of Section~III-C to these outputs yields $2.9$~bit/symbol at a \ac{kmr} of $0.04$ without Eve, and $0.8$~bit/symbol at $0.10$ under the perfect-position Eve, so the pipeline runs end to end on the synthesized data.

For the real-world datasets, $164$ and $172$ measurement pairs, we pretrain the framework on the much larger synthesized dataset and fine-tune on the real-world measurements with the Transformer backbone frozen and only the final linear projection retrained. This single-layer adaptation limits overfitting and tests transfer of the pretrained encoder. Eve's predictor is fine-tuned in full to keep the adversarial bound tight.

Table~\ref{tab:s4pls-meas} reports the \ac{vpq} stage of the fine-tuned models. The $95\%$ confidence intervals follow from the delta method\cite{vandervaart1998}, since every reported quantity is a smooth function of sample means over the test split; the same construction over Table~\ref{tab:s4pls-syn}'s $10\,209$-sample split gives half-widths of at most $0.02$ on every quantity. Without fine-tuning the synthesized encoder transfers poorly: $\Pr\{W=V\}$ is $0.216$ in the lab room and $0.388$ in the chamber, and below $0.09$ under a correlated Eve. After fine-tuning the no-Eve agreement reaches $0.780$ and $0.823$. The chamber outperforms the lab room when Eve is absent or uncorrelated, reflecting its cleaner scattering environment. Under an uncorrelated Eve the source is independent of $Z$ by construction, and the confidence interval on $\calI_{\mathrm{VUB}}$ contains zero at both sites; under a correlated Eve it is strictly positive at every operating point. The maximal-leakage column moves with $\calI_{\mathrm{VUB}}$, $1.7$ to $5.6$ times above the uncorrelated-Eve value at the same site. Each row is a separately fine-tuned encoder: against the perfect-position Eve the fine-tune buys secrecy with agreement, so those rows carry the lowest leakage and the lowest agreement at once. Under the perfect-position Eve the agreement falls to $0.112$ and $0.186$, that is, $p_e=0.888$ and $0.814$; at these mismatch rates Theorem~\ref{thm:universal-rate} gives $R^{\star}_{\mathrm{univ}}=(1-h_q(p_e))\log|\calW|$ of $0.025$ and $0.127$~bit/symbol at $|\calW|=16$.

%% file: paper-sections/conclusion.tex
In this work, we introduced a variational \ac{cr} extraction framework, consisting of two stages. In the first stage, \ac{vpq} learns probabilistic encoders that quantize correlated observations at Alice and Bob into nearly uniform and highly correlated \acp{rv} while suppressing information leakage to Eve via an adversarial mutual information objective.
In the second stage, a $q$-ary linear code-offset secure sketch reconciles the quantized outputs into identical secret keys, with perfect secrecy under zero leakage and an $\epsilon$-bounded leakage rate when the residual leakage is nonzero (recovering weak secrecy as $\epsilon\to 0$).

We first evaluated the framework on fading channels against one classical and two recent learning-based baselines across the full \ac{snr} range: the outputs reached near-maximal entropy and high agreement, and under a correlated Eve the framework leaked the least of all methods.
We then applied the framework to sensing-based \ac{plk} generation in \ac{isac} systems, where \ac{ra} maps serve as the source of \ac{cr}. End-to-end 5G NR simulations delivered reconciled keys under every eavesdropper considered, and real-world \ac{sdr} measurements confirmed that the learned quantizer transfers to hardware, while a transfer-learning design with frozen backbone and last-layer fine-tuning lets pretrained models generalize across environments. Robustness was further tested across a grid of correlated eavesdroppers with controllable position uncertainty, down to a perfect-position attacker who observes Bob's exact range and angle. Compared with conventional \ac{plk} schemes that rely on reciprocity and require two-way channel probing, our method reduces protocol overhead, supports high-mobility scenarios, and integrates secrecy against a passive position-aware eavesdropper without a separate privacy-amplification step.

The \ac{vpq} training objectives were shown to translate directly into lower bounds on the one-way secret-key capacity of the induced discrete source, with the universal-rate benchmark $(1 - h_q(p_e))\log q$ as the optimal one-way key rate achievable uniformly over the sources that share the output's alphabet size and mismatch probability. Several directions follow from this work. One is to bring the public reconciliation message inside the training objective. Another is to extend the adversary from a passive to an active eavesdropper. The sensing source itself can be widened: a single-symbol \ac{ra} map discards the Doppler dimension, and reciprocity-based channel probing would add a second correlated observation. Larger measurement campaigns, across more environments and mobile trajectories, would extend the validation.